\documentclass[titlepage,12pt]{utarticle}

\usepackage{amsmath}
\usepackage{mathrsfs}
\long\def\omit#1{}

\newcommand\HH{\mathcal{H}}

\newcommand\MM{\mathcal{M}}

\newcommand\OO{\mathcal{O}}
\newcommand\DD{\mathcal{D}}

\newcommand\dM{\partial \MM}

\newcommand\eps{\epsilon}

\newcommand\nts{\negthickspace}
\newcommand\bns{\nts \nts \nts}

\newcommand\tg{\tilde{g}}

\newcommand\de{\delta}
\newcommand\vf{\varphi}
\newcommand\vfb{\bar{\varphi}}
\newcommand\ve{\varepsilon}

\newcommand\pvf{\partial_{\vf}}
\newcommand\ups{\upsilon}

\DeclareMathAlphabet{\mathpzc}{OT1}{pzc}{m}{it}
\newcommand\Ham{\mathscr{H}}
\newcommand\Lag{\mathscr{L}}

\begin{document}

\preprint{MCTP--04-06\\ {\tt hep-th/0402050}\\ }

\title{Holography, Diffeomorphisms, and Scaling Violations in the CMB}

\author{Finn Larsen\footnote{\texttt{larsenf@umich.edu}} ~and Robert McNees\footnote{\texttt{ramcnees@umich.edu}}}

\oneaddress{Michigan Center for Theoretical Physics\\
  University of Michigan\\
        Ann Arbor, MI-48109, USA  }

\date{}

\Abstract{ 
We analyze diffeomorphism invariance in inflationary spacetimes regulated 
by a boundary at late time. We present the action for quadratic fluctuations 
in the presence of a boundary, and verify that it is gauge invariant precisely
when the correct local counterterms are included. The scaling behavior of bulk 
correlation functions at the boundary is determined by Callan-Symanzik equations 
which predict scaling violations in agreement with the standard inflationary 
predictions for spectral indices of the CMB.
}

\maketitle

\section{Introduction} 
\label{sec:introduction}
It is useful to think of cosmological evolution in terms of a family of spatial slices,
with time appearing as a parameter identifying the individual slices. 
The cosmological wave function \cite{Hartle:1983ai} is then a functional which, in 
the Schr\"{o}dinger picture, takes the schematic form:
\begin{equation}
\Psi[\phi] \sim e^{iS[\phi]}
\label{cosmowave}
\end{equation}
where $\phi$ is a set of variables defined on a three dimensional equal-time slice. In the 
semiclassical approximation $S[\phi]$ can be identified as the Hamilton-Jacobi (H-J) 
functional, a versatile tool in 
cosmology \cite{Salopek:1990jq,Salopek:1994sq,Salopek:1997he,Parry:mw}. 
The H-J functional is defined as the on-shell action, interpreted as a functional of the dynamical variables on the equal time slice; so the H-J form of the dynamical problem involves gravitational 
physics on a `bulk' manifold which ends at a `boundary', the equal time slice under 
consideration. It is therefore well suited to the study of gravitational physics on manifolds with a boundary, a problem which also has many other applications, such as in 
brane-world models. 

In the present paper we study the gravity-scalar system on a manifold which, for definiteness,
we take as an inflationary spacetime. The on-shell actions for these spacetimes contain late 
time divergences which can be regulated by truncating the manifold at a late time, resulting in a 
boundary. As a result the action in \eqref{cosmowave} is a functional of the `boundary data', 
the variables $\phi$ evaluated on the spatial slice corresponding to the late time cut-off.
Our main results are:
\begin{itemize}
\item
Diffeomorphism Invariance is not automatic in the presence of such a boundary. The simplest way 
to preserve diffeomorphism invariance is to introduce local counterterms on the
boundary. We determine their form. 
\item
We compute the quadratic action for fluctuations around a manifold with a
boundary. We present our result in terms of gauge invariant variables. 
\item
We interpret the spectral indices of the Cosmic Microwave Background (CMB)
in terms of scaling violations of a `boundary theory'. This perspective is 
holographic in character, since a three-dimensional theory controls the 
four-dimensional physics.
\end{itemize}

The starting point for our discussion is the straightforward and explicit computation 
of the H-J functional~\cite{Maldacena:2002vr,Larsen:2003pf}. 
The result of this computation suffers from a divergence as 
the time of the slice is taken to future infinity. This divergence is dominated by 
large wavelengths, and so it can be cancelled by adding a local boundary term --- 
a counterterm --- to the action. 

However, the na\"{i}ve computation suffers from additional, and seemingly
more serious, problems. As we will explain, the presence of an arbitrary boundary, 
introduced to regulate divergences, renders the H-J functional inconsistent with the
full set of four dimensional diffemorphisms.
This failure of local reparametrization invariance can be remedied by
supplementing the standard action with a boundary term. We will show that 
this boundary term, designed to restore diffeomorphism invariance, is 
in fact the same as the counterterm needed to cancel infrared divergences. 

The gravity side of the AdS/CFT correspondence \cite{Maldacena:1997re,Gubser:1998bc,Aharony:1999ti}
involves the on-shell action on a (possibly deformed) AdS-space. It is well known that this action exhibits
infrared divergences due to the behavior of the metric near the boundary at spatial infinity. These divergences
are naturally cancelled by the introduction of boundary counterterms \cite{Witten:1998qj,Balasubramanian:1999re,Emparan:1999pm,Kraus:1999di}.
In the context of the AdS/CFT correspondence,
the counterterms are interpreted in the dual conformal field theory as the usual 
counterterms needed to cancel ultraviolet divergences in quantum field 
theory \cite{Balasubramanian:1999re};
but their origin on the gravity side is less clear. 
Although we work with cosmological spacetimes for definiteness, our results are valid for asymptotically AdS-spaces as well.
This suggests a new perspective on the counterterms in AdS/CFT, which is 
rooted solidly in gravity: counterterms are needed to maintain diffeomorphism 
invariance. 


In standard cosmological perturbation theory it is customary to implement 
diffeomorphism invariance by introducing gauge invariant physical observables
\cite{Lifshitz:em,Bardeen:kt,Mukhanov:jd,Mukhanov:1990me,Riotto:2002yw}. 
We will extend this result to include the boundary, and present the quadratic action 
for fluctuations in this more general case, again written in gauge invariant form. 
This is one of our main results.

Diffeomophism invariance also constrains the dependence of the boundary 
theory on the gauge invariant variables. The origin of these additional constraints 
are the diffeomorphisms acting on the direction normal to the boundary. These 
transformations are implemented as scaling symmetries on the three-dimensional 
equal time surface, and so their effect is to determine the scale dependence of the 
correlation functions.  We refer to the equations determining the scale 
dependence as the Callan-Symanzik equations. 

The Callan-Symanzik equations can be solved using techniques that are standard from 
renormalization group theory. The result of this analysis is general formulae for the 
correlation functions of the theory, determined by symmetries alone. To exploit these 
formulae, one must add dynamical input, {\it e.g.} from slow roll inflation. Given this input, 
our expressions are renormalization group improved versions of the more conventional 
results. We consider both scalar and tensor fluctuations and determine, in particular,
the scalar and tensor mode spectral indices, $n_s$ and $n_t$, which characterize 
the scale dependence of the CMB. 

The terminology introduced to describe consequences of diffeomorphism 
invariance --- counterterm, Callan-Symanzik equations, and the Ward identity --- 
is  that of a local quantum field theory on the equal time slice. In the context
of the AdS/CFT correspondence our terminology fully justified but, in cosmology, it 
refers to a conjectured dS/CFT correspondence \cite{Strominger:2001pn, Strominger:2001gp, Klemm:2001ea, Spradlin:2001nb, Witten:2001kn, Balasubramanian:zh}. Such ideas, though rather speculative, have been used to address inflation \cite{Maldacena:2002vr, Larsen:2003pf, Larsen:2002et, vanderSchaar:2003sz}.
It would be extremely interesting if a truly 
holographic theory of cosmology could be established. However, whether cosmological
holography is true or not, the counterterms we discuss are a universal part of the 
gravitational action, determined from 
diffeomorphism invariance alone. It may be that gravity is characterized by 
``infrared universality classes" which would be similar to the ``ultraviolet 
universality classes" familiar from quantum field theory. In quantum field
theory truly short distances decouple from long distance physics and similarly
it seems that, in cosmology, truly large distance physics, beyond the horizon, 
decouples from short distance physics, {\it i.e.} observable cosmology \cite{Schalm:2004qk}. The 
notion of ``infrared universality classes" could develop into a framework 
for addressing the notorious fine-tuning problems in cosmology. This might apply not only
to the fine-tuning problems normally associated with inflation, but also to other naturalness
problems associated with the decoupling of long and short distance physics, such as the
cosmological constant problem.

This paper is organized as follows. 
In section 2 we review the cancelling of infrared divergences {\it via} the introduction 
of counterterms. We then present an argument that identifies the origin of the 
counterterms as diffeomorphism invariance. In section 3 we compute the quadratic 
action of fluctuations around the background.
To do this, we review the standard notion of gauge invariance in the bulk theory, 
and show how this can be extended to the boundary, precisely when the correct 
boundary terms are introduced. In Section 4 we discuss the consequences of 
diffeomorphism invariance for the form of correlation function. The constraints
are summarized by a master equations which, in a special case, reduces to
the Callan-Symanzik equation. 
In section 5 we solve the Callan-Symanzik equation. In particular we determine
the spectral parameters of cosmological inflation as the scaling violations of the 
theory. 
Finally, in section 6, we conclude with an outlook for further developments.

\section{Counterterms on the Boundary}
\label{bndyctterms}
In this section we review the appearance of infrared divergences in the gravitational action 
and their cancellation by counterterms on the regulating boundary. We then discuss how 
the presence of the boundary introduces sources in the equations of motion and, by a 
related mechanism, violates diffeomorphism invariance. This shows that  
counterterms are needed to preserve a crucial symmetry, diffeomorphism 
invariance. 

\subsection{The Setting} 
\label{subsec:review}
The simplest and most common setting for discussing inflation is
four-dimensional Einstein gravity coupled to a single scalar field. 
For a review of scalar field inflation, see \cite{Lyth:1998xn,Peacock:ye,Peebles:xt, Liddle:cg} and references therein. The action of the theory is:
\begin{eqnarray} \label{inflationaction}
   S & = & \int_{\MM} \bns d^4 x \sqrt{g} \left( \frac{1}{16 \pi G} \, R - \frac{1}{2} \, \nabla^{\mu} \vf \nabla_{\mu} \vf - V(\vf) \right)
		- \frac{1}{8\pi G} \, \int_{\dM} \bns d^3 x \sqrt{\tg} \,K
\end{eqnarray}
The cosmological spacetime $\MM$ will have a spacelike boundary $\dM$ 
defined by a timelike unit normal $n^{\mu}$. The metric on $\MM$ is 
$g_{\mu\nu}$ and the induced metric on $\dM$ is $\tg_{\mu\nu}$. The Gibbons-Hawking 
term \cite{Gibbons:1976ue} ensures that the action poses a well defined variational problem.
In the rest of this paper we use units where $8 \pi G = 1$.

During the inflationary epoch the metric and the scalar field are approximately
spatially homogeneous. They take the form:
\begin{eqnarray} \label{gravscalarsystem}
	\vf(\vec{x},\tau) & = & \vf(\tau) + \chi(\vec{x},\tau) \\ \nonumber
	g_{\mu\nu}(\vec{x},\tau) & = & a(\tau)^2 \, \eta_{\mu\nu} + h_{\mu\nu} (\vec{x},\tau)
\end{eqnarray}
where $\eta_{\mu\nu} = \mathrm{diag}\{-1,1,1,1\}$ is the usual Minkowski metric. The 
$\vec{x}$ are spatial coordinates and $\tau$ is conformal time, which runs over 
$\tau \in (-\infty,0)$. Standard slow-roll inflation assumes that $\varphi$ is approximately 
constant, corresponding to a background that is approximately de Sitter space, 
{\it i.e.} $a(\tau)\propto\tau^{-1}$. These `quasi-de Sitter' spacetimes will be our
primary example but most of our results in fact apply to general FRW cosmologies. 

We refer to the spatially homogenous parts of \eqref{gravscalarsystem} as the ``background". 
The equations of motion governing the background are the standard FRW equations: 
\begin{eqnarray}\label{scalarEOM}
	&~& \vf\,'' + 2 \HH \vf\,' + a^2 \pvf V  =  0 \\
\label{metricEOM}
	&~& 3 \, \left(\frac{\HH}{a}\right)^2  =  \frac{1}{2} \, \left(\frac{\vf\,'}{a}\right)^2 + V
\end{eqnarray}
A prime denotes a derivative with respect to the conformal time $\tau$, and $\HH = a' / a$ is the conformal Hubble factor. 

The dynamics of the small, spatially inhomogenous fluctuations in \eqref{gravscalarsystem} 
is described by the action \eqref{inflationaction},  evaluated as a series in $\chi$ and 
$h_{\mu\nu}$ around the background. Terms in the action which are linear in the 
fluctuations are ``first order", and quadratic terms are ``second order". According to
inflation \cite{Starobinsky:ty, Starobinsky:ee, Starobinsky:83, Mukhanov:xt, Mukhanov:nu, Guth:ec, Hawking:1982cz, Bardeen:qw}
these fluctuations 
are responsible for the minute variations in the CMB that we observe today \cite{Bennett:2003bz, Hinshaw:2003ex, Peiris:2003ff}. 

\subsection{Counterterms: Cancelling Divergences}  
A powerful way to analyze a dynamical system is the Hamilton-Jacobi (H-J) formalism. 
The linchpin of this formalism is the H-J functional, defined as the on-shell 
action:
\begin{equation}
S_{\rm H-J}[\varphi,\tg_{\mu\nu}] = S_{\rm on~shell}[\varphi,\tg_{\mu\nu}] 
\end{equation}
The dynamical variables $\varphi$ and $\tg_{\mu\nu}$  are defined on a spatial 
slice parametrized by the conformal time $\tau$. The initial conditions are
usually left implicit but, in the cosmological context, it is natural to specify 
them by imposing 
regularity as $\tau\to-\infty$. The H-J functional can be interpreted in the 
semiclassical regime as the phase of the cosmological wave 
function \eqref{cosmowave}. The regularity conditions at early times then specifies 
the initial conditions as the Hartle-Hawking state. 

The H-J functional corresponding to the action  \eqref{inflationaction} diverges 
as the time parameter is taken to future infinity $\tau\to 0^-$. This can be seen 
in an elementary way by making simplifying assumptions about
the background and the fluctuations. Consider, for example, a homogeneous
massless scalar field in the background of pure de Sitter space. The bulk equations
of motion then determine the scalar field as a linear combination of decreasing 
and increasing solutions, with the regularity condition at early times excluding
the ``decreasing" solution. However, the on-shell action, evaluated on the ``increasing"
solution, diverges at late times. As a result the H-J functional suffers an infrared 
divergence (for details see \cite{Larsen:2003pf}). 
In this elementary derivation the origin of the infrared divergence is that,
in general, no solution gives regular on-shell actions at both early and late times.

The divergences can be characterized much more generally by considering the 
Hamiltonian constraint which, in the present context, is implemented by 
the H-J equation:
\begin{equation}
\left( \frac{1}{\sqrt{\tg}} \tg_{\mu\nu} \frac{\delta S}{\delta \tg_{\mu\nu}}\right)^2
-2\left( \frac{1}{\sqrt{\tg}}\frac{\delta S}{\delta \tg_{\mu\nu}} \right)
\left( \frac{1}{\sqrt{\tg}}\frac{\delta S}{\delta \tg^{\mu\nu}}\right)
= V - \frac{1}{2}{\cal R} + \frac{1}{2}\vec{D}\varphi\cdot\vec{D}\varphi
\label{HJequation}
\end{equation}
The dominant terms in the solutions are captured by the inverse metric 
expansion:
\begin{eqnarray} 
\label{caction}
	S = \int_{\dM} \bns d^3 x \sqrt{\tg} \left( U(\vf) + 
	M(\vf) \vec{D} \vf \cdot \vec{D}\vf + C(\vf) \tilde{R} +\ldots \right)
\end{eqnarray}
In this form the divergence of the H-J functional is a consequence of 
the divergence of the FRW scale factor. The terms displayed explicitly
in the {\it ansatz} \eqref{caction} are sufficient to capture {\it all} divergences in the 
action, even when the terms indicated by dots are neglected. This characterization 
of the divergences in the action is rather general because it applies to the full 
cosmology, with inhomogeneous fluctuations, and it makes only modest 
assumptions about the background\footnote{The conformal factor must diverge 
at late times. This is satisfied for all expanding cosmologies. In addition, 
we use the de Sitter form of the metric to estimate when the 
series \eqref{caction} can be truncated. This is satisfied for 
``quasi-de-Sitter spacetimes",  such as those that appear in inflationary 
cosmology.}. It should also be noted that this type of local ansatz for the on-shell action is commonly used in the study of holographic renormalization group flows in AdS/CFT \cite{deBoer:1999xf, Bianchi:2001kw, Bianchi:2001de, DeWolfe:2000xi, Balasubramanian:1999jd, Freedman:1999gp, Nojiri:2002uz}.

Inserting the {\it ansatz} \eqref{caction} in the H-J equation, and solving order-by-order in the 
inverse metric expansion, we find differential equations that determine the
functions $U(\varphi)$, $M(\varphi)$, and $C(\varphi)$:
\begin{eqnarray} \label{Ueqn}
	0&=&\frac{3}{4} \, U^2 - \frac{1}{2} \left( \pvf U\right)^2 - V   \\ \label{Ceqn}
	0&=&  1 + U \, C - 2 \, \pvf U \, \pvf C   \\ \label{Meqn}
	0 &  = & 1 - U\,M - 4\pvf U \,\pvf C + 2 \, \pvf U \, \pvf M + 4\pvf^{\,2}\,U\,M      
\end{eqnarray}
The first equation \eqref{Ueqn} reproduces the Friedmann equation \eqref{metricEOM}
for a homogenous background if we identify $U(\vf)$ and its first
derivative as:
\begin{equation} 
U(\varphi)=-2\,\frac{{\cal H(\varphi)}}{a(\varphi)}\,\,\,\,\,\,\,\,\pvf U = \frac{\vf\,'}{a}
\label{Ures}
\end{equation}
It is important to note that the equation \eqref{Ueqn} and the conditions \eqref{Ures}
specify the functional dependence of $U(\vf)$ on the field $\vf$, and not just the value
that $U(\vf)$ and its first derivative take. It may therefore be used as the leading counterterm,
even for backgrounds which contain spatial inhomogeneities.

At this point we have shown that the H-J functional $S$ diverges, and that the
divergences are characterized in general by \eqref{caction}. We can use this
information to isolate the finite part of the H-J functional as follows: first, regulate
the divergences that appear in the action by cutting the spacetime off at some late
time $\tau_0$. Then introduce the following counterterms, intrinsic to the spatial
hypersurface defined by the condition $\tau=\tau_0$.
\begin{eqnarray} 
\label{ctaction}
	S_{\rm CT}(\tau_0)  = \int_{\dM_0} \bns d^3 x \sqrt{\tg} \left( U(\vf) + 
	M(\vf) \vec{D} \vf \cdot \vec{D}\vf + C(\vf) \tilde{R} \right)
\end{eqnarray}
The `total action' $S_{\rm tot}$ is then given by the original action, minus the contribution
of the counterterms on the regulating boundary:
\begin{equation}
	S_{\rm tot}(\tau_0)  =  S (\tau_0) - S_{\rm CT}(\tau_0) 
	\label{totact}
\end{equation}
This action $S_{\rm tot}$ is finite as $\tau_0\to 0$. 
A simple physical interpretation of this procedure is that $S$ is the action 
of the complete cosmology, $S_{\rm CT}$ captures 
the divergences present in the background, and their difference, $S_{\rm tot}$, 
represents the action of the fluctuations alone. This interpretation should
not be taken too seriously: since
$S_{\rm CT}$ is a true functional of the spacetime and so, 
when expanding around a background, there will be terms attributable 
to the fluctuations. A more appropriate, albeit more abstract, terminology 
is that of standard quantum field theory: $\tau_0$ is the {\it physical cut-off}, 
\eqref{ctaction} are the local {\it counterterms}, and \eqref{totact} is the 
{\it renormalized action}. 

\subsection{Sources at the Boundary}
\label{sourcebndy}
The counterterms are intrinsic to the regulating boundary, so they do not affect the 
{\it bulk} equations of motion derived from \eqref{inflationaction}. However, we need 
to consider the equations of motion {\it on} the boundary. The equations of motion
are determined by the variational principle. The first order variation of the original
action \eqref{inflationaction} is:
\begin{eqnarray}
\label{deltaS}
	\delta S & = & \int_{\MM} \bns d^4 x \, \sqrt{g} \, \left( \frac{1}{2} \, h^{\mu\nu} 
	\left( G_{\mu\nu}-T_{\mu\nu}\right) + \chi \, 
			\left( \nabla^2 \, \vf - \pvf V \right)\right) \\ \nonumber
&   & + \int_{\dM} \bns d^3 x \sqrt{\tg} \, \left( h^{\mu\nu} \, \pi_{\mu\nu} + \chi \, \pi_{\vf} \right) 
\end{eqnarray}
In the usual variational principle we consider only variations of the fields that vanish on 
the boundary $\dM$; and so we discard the boundary term in \eqref{deltaS}. 
Extremizing the action then gives the Einstein equations $G_{\mu\nu} = T_{\mu\nu}$ and the 
scalar equation of motion $\nabla^2 \vf = \pvf V$, as expected.  

The equations of motion {\it on} the regulating boundary  with conformal time $\tau=\tau_0$
are determined by keeping the variations $h_{\mu\nu}$ and $\chi$ \emph{arbitrary} there. We
see that: 
\begin{equation}
	\pi_{\vf} \, = \, \frac{1}{\sqrt{\tg}} \, \frac{\delta S}{\delta \vf} ~~~~~~~
	\pi^{\mu\nu} \, = \, \frac{1}{\sqrt{\tg}} \, \frac{\delta S}{\delta \tg_{\mu\nu}}
\end{equation}
correspond to sources on the boundary which act as net forces. 
The sources $\pi_{\mu\nu}$ and $\pi_{\vf}$ are the canonical momenta which, in general, 
are non-vanishing. The action is not extremized, $\delta S \neq 0$, unless these terms
are cancelled by external forces.

In the present context the appropriate action is in fact $S_{\rm tot}=S -S_{\rm CT}$, rather than
just S, and the boundary counterterms provide an external force. The momenta 
due to the counterterms are:
\begin{eqnarray}
\label{Pdefs}
	P_{\vf} \, = \, \frac{1}{\sqrt{\tg}} \, \frac{\delta S_{CT}}{\delta \vf} & & 
	P^{\mu\nu} \, = \, \frac{1}{\sqrt{\tg}} \, \frac{\delta S_{CT}}{\delta \tg_{\mu\nu}}
\end{eqnarray}
so the conditions for the variation of $S_{\rm tot}$ to vanish on the boundary are:
\begin{equation}
\pi_{\vf}=P_{\vf}~~~~~~~ \pi^{\mu\nu} = P^{\mu\nu}
\label{consieqs}
\end{equation}
It is not easy to solve the equations \eqref{consieqs} in general. In the special case 
of a homogeneous background it is straightforward to compute the canonical momenta in 
terms of the scale factor $a(\tau)$ and its derivatives, and then find the counterterm momenta 
with appropriate variations. The result is:
\begin{equation}
	U(\varphi) = -2 \,\frac{\HH(\varphi)}{a(\varphi)} ~~~~~~~~ \pvf U = \frac{\vf\,'}{a} 
\end{equation}
This agrees precisely with the result \eqref{Ures} for the counterterms found by expanding
the H-J equation and identifying divergent terms. 

In some contexts it is a consistency requirement that the equations of motion are
satisfied, even on the boundary. Thus it is possible to reverse the logic and take
the equations of motion in the boundary as the starting point which motivates
the introduction of the counterterms in the first place. From this point of
view the infrared divergences and, in particular, the limit $\tau_0\to 0^-$
play a secondary role. However, we will not pursue this in detail, but rather go 
one step further and motivate the counterterms from symmetries that hold even when 
the equations of motion do not. This is the subject we turn to next. 

\subsection{Counterterms: Diffeomorphism Invariance}
Perhaps the best way to motivate boundary counterterms is that they are required 
to maintain diffeomorphism invariance on spacetimes with a boundary. This 
subsection presents the argument.   

Diffeomorphism invariance states that the physics of a system is independent of the 
coordinate system chosen to describe it. Stated as a local symmetry, it means that the 
action $S$ of a theory containing gravity is invariant under the infinitesimal local change 
of coordinates $x^{\mu} \rightarrow x^{\mu}+\eps^{\mu} (x)$. Under such a coordinate 
change the metric transforms as:
\begin{eqnarray}
\label{mettr}
  g_{\mu\nu} & \rightarrow & g_{\mu\nu} - \nabla_{\mu} \eps_{\nu} - \nabla_{\nu} \eps_{\mu} 
\end{eqnarray}
and the scalar $\vf$ transforms as:
\begin{eqnarray}
	\vf & \rightarrow & \vf - \eps^{\mu} \nabla_{\mu} \vf
	\label{scalartr}
\end{eqnarray}
The variation of the action \eqref{inflationaction}
under an infinitesimal diffeomorphism is:
\begin{eqnarray}
	\delta_{\eps} S & = & \int_{\MM} \bns d^4 x \sqrt{g} \left[ \nabla^{\mu} \eps^{\nu}
		 \, \left( G_{\mu\nu} - T_{\mu\nu} \right) - \eps^{\mu} \nabla_{\mu} \vf \,
		\left( \nabla^{2} \vf - \pvf V\right)\right] \\ \nonumber
	& & + \int_{\dM} \bns d^3 x \sqrt{\tg} \left[ -2  \pi_{\mu\nu} \, \nabla^{\mu} \eps^{\nu} 
		- \pi_{\vf} \, \eps^{\mu} \nabla_{\mu} \vf \raisebox{10pt}{\,}\right]
\end{eqnarray}
After integration by parts this becomes:
\begin{eqnarray}\label{intermediateresult}
	\delta_{\eps} S & = & \int_{\MM} \bns d^4 x \sqrt{g} \left[ - \eps^{\nu} \nabla^{\mu} G_{\mu\nu}
	 + \eps^{\nu} \left(\nabla^{\mu} T_{\mu\nu} -  \nabla_{\nu} \vf \left( \nabla^{2} \vf 
	 - \pvf V\right)\right) \right] \\ \nonumber
	& & + \int_{\dM} \bns d^3 x \sqrt{\tg} \left[ n^{\mu} \eps^{\nu} \left( G_{\mu\nu} - 
	T_{\mu\nu}\right) -2  \pi_{\mu\nu} \, \nabla^{\mu} \eps^{\nu} 
	- \pi_{\vf} \, \eps^{\mu} \nabla_{\mu} \vf \raisebox{10pt}{\,} \right] 
\end{eqnarray}
The contracted Bianchi identity implies $\nabla^{\mu} G_{\mu\nu} = 0$ and the remaining
bulk terms vanish as well\footnote{To verify that $\nabla^{\mu} T_{\mu\nu}=\nabla_\mu\varphi (\nabla^2\varphi -\partial_\varphi V)$ simply insert 
the explicit form of the energy-momentum tensor 
$T_{\mu\nu}=\nabla_\mu\varphi\nabla_\nu\varphi-g_{\mu\nu}(\frac{1}{2}(\nabla\varphi)^2+V(\varphi))$.}. The fact that {\it all} bulk terms vanish identically is independent of the equations
of motion, which have not been imposed.

Variations of the action under diffeomorphisms are thus captured
by certain boundary terms. These 
terms can be written in a more illuminating form by splitting the 
vector $\eps^{\mu}$ into its normal and tangential components:
\begin{eqnarray}
	\eps^{\mu} & = & -n^{\mu} n_{\nu} \eps^{\nu} + \tg^{\mu}_{~\nu} \, \eps^{\nu}
\end{eqnarray}
It can be shown that the Gauss-Codazzi equations and the definitions of canonical 
momenta imply:
\begin{eqnarray}
n^\mu n^\nu (T_{\mu\nu}-G_{\mu\nu}) &=& 
2\pi^{ij}\pi_{ij}-\pi^i_i\pi^j_j - \frac{1}{2}{\cal R} + \frac{1}{2}\pi^2_\varphi + 
\frac{1}{2}\vec{D}\varphi\cdot\vec{D}\varphi + V = \Ham \\
n^\mu \epsilon^i (G_{\mu i} - T_{\mu i}) &=& \pi_\varphi ~\epsilon^i D_i\varphi
-2\epsilon^i D^j \pi_{ij} = -\epsilon^i \Ham_i
\end{eqnarray}
and so we can write \eqref{intermediateresult} as:
\begin{eqnarray}\label{finalresult}
 \delta_{\eps} S & = & \int_{\dM} \bns d^3 x \sqrt{\tg} \left[ n_{\lambda} \eps^{\lambda} \, \Ham - 
	\eps^{i} \Ham_{i} - 2 \pi^{\mu\nu} \nabla_{\mu} \eps_{\nu} - \pi_{\vf} \eps^{\mu}
	\nabla_{\mu} \vf\right]
\end{eqnarray}
This makes it manifest that the only components of the equations of motion 
$G_{\mu\nu}-T_{\mu\nu}=0$ which appear as generators of the diffeomorphisms 
are those proportional to the contraints $\Ham=\Ham_i=0$. 

Splitting the remaining terms into normal and tangential components gives:
\begin{eqnarray}
2\pi^{\mu\nu}\nabla_\mu\epsilon_\nu &=& 2 \pi^{ij}D_i\epsilon_j + 4(\pi^{ij}\pi_{ij}
-\frac{1}{2}\pi^i_i \pi^j_j) n_\lambda\epsilon^\lambda \\
\pi_\varphi ~\epsilon^\mu \nabla_\mu\varphi &=& \pi_\varphi \epsilon^i D_i\varphi
+\pi^2_\varphi ~n_\lambda \epsilon^\lambda
\end{eqnarray}
and the complete expression can be simplified as:
\begin{equation}
\label{diffeoS}
	\delta_{\eps} S  =  - \int_{\dM} \bns d^3 x \sqrt{\tg} \, n_{\lambda} \eps^{\lambda} \, \Lag
\end{equation}
In other words, the Lagrangian density 
transforms as a scalar field \eqref{scalartr}, as it should\footnote{The Lagrangian $\Lag$
that appears here is in first order form, {\it i.e.} the Gibbons-Hawking term has been 
cancelled through integration by parts.}. The result \eqref{diffeoS}
shows that the action is invariant under reparameterizations of the spatial coordinates
but, when there is a boundary present, the action is {\it not} invariant under reparametrizations 
of the direction normal to the boundary. One often neglects this variation by assuming that the 
coordinate transformations are localized in the bulk, so that the normal component of 
$\eps^{\mu}$ falls off sufficiently rapidly near the boundary $\dM$. In the present context we are 
not entitled to ignore this term. Indeed it is of great interest, because it characterizes the violation 
of diffeomorphism invariance due to the non-covariant regulator we have introduced.

So far we have neglected the counterterms in the discussion of diffeomorphism invariance. 
The variation of the renormalized action $S_{\rm tot}=S-S_{\rm CT}$ under an infinitesimal 
diffeomorphism follows from \eqref{finalresult} and the definitions \eqref{Pdefs}. It can be
written as:
\begin{eqnarray}
	\delta_{\eps} S_{\rm tot} & = & 
	\int_{\dM} \bns d^3 x \sqrt{\tg} \left[ n_{\lambda} \eps^{\lambda} \, \Ham - 
	\eps^{i} \Ham_{i} - 2 \left( \pi^{\mu\nu} - P^{\mu\nu} \right) \nabla_{\mu} \eps_{\nu} - 
	 \left( \pi_{\vf} - P_{\vf} \right) \eps^{\mu} \nabla_{\mu} \vf \raisebox{10pt}{\,}\right]
	\label{fullvar}
\end{eqnarray}
Invariance  under the full four-dimensional diffeomorphisms requires that this expression vanishes.
It is natural to impose the usual Hamiltonian and momentum constraints:
\begin{eqnarray}
	\Ham \, = \, 0 & & \Ham_i \, = \, 0
\end{eqnarray}
The remaining terms in \eqref{fullvar} are then similar to the boundary sources for momentum, 
discussed in subsection \ref{sourcebndy}. However, here we have not used the equations of 
motion; so these terms constitute genuine violations of diffeomorphism invariance. 

The point, of course, is that the counterterms, when appropriately chosen, can restore 
diffeomorphism invariance; indeed, this might be the most persuasive motivation for 
introducing the counterterms in the first place.
The condition for diffeomorphism invariance in the presence of a boundary
is:
\begin{eqnarray}\label{diffeomomentumcond}
	\pi_{\mu\nu} \, = \, P_{\mu\nu} & & \pi_{\vf} \, = \, P_{\vf}
\end{eqnarray}
as in subsection \ref{sourcebndy}. As explained there, these 
equations can be integrated,
for homogeneous backgrounds, to determine the counter-terms as:
\begin{eqnarray}
	U(\varphi) \, = \, - 2 \, \frac{\HH(\varphi)}{a(\varphi)} & & \pvf U \, = \, \frac{\vf \, '}{a}
	\label{Uctdiff}
\end{eqnarray}
Note that the equalities in \eqref{diffeomomentumcond} relate the values that the momenta take at the boundary, and not their functional form. This can be clearly seen when one recalls that the
canonical momenta $\pi_{ij}$ and $\pi_{\vf}$ depend on the normal derivatives of the boundary
data, whereas the counterterms, and hence their contribution to the momenta, are entirely intrinsic to the boundary.

In section \ref{sec:DiffInv} we will consider inhomogeneous backgrounds as well, and 
show similar agreements for the subleading counterterms $M(\varphi)$ and 
$C(\varphi)$. Diffeomorphism invariance is therefore equivalent to
the condition that divergences cancel at late times. 

\section{Gauge Invariance and the Quadratic Action} \label{sec:DiffInv}
The purpose of this section is to derive the quadratic action for fluctuations in the
presence of a boundary, and demonstrate that it is gauge invariant. We first review gauge 
invariance for bulk fluctuations. We then include boundaries, and present the
action in this case. We show that this action can be written in terms of gauge invariant 
variables precisely when the correct counterterms have been included. 

\subsection{Gauge Invariance for Inflationary Spacetimes}
\label{subsec:gaugeinv}
We consider a spatially homogenous background defined by a scalar field $\vf(\tau)$
and a scale factor $a(\tau)$. The fluctuations correspond to the terms $\chi$ and $h_{\mu\nu}$ appearing in \eqref{gravscalarsystem}, with $h_{\mu\nu}$ parameterized 
by\footnote{We consider only scalar and tensor fluctuations in the metric. Vector 
fluctuations do not couple to energy density perturbations in the cases we are 
interested in.}:
\begin{eqnarray} \label{scalarfluctuations}
	h_{\mu\nu} & = & a(\tau)^2 \left(\begin{matrix}
	2\,\Phi(\vec{x},\tau) & \partial_i B(\vec{x},\tau) \\ 
         \partial_i B(\vec{x},\tau) & \phi_{ij}(\vec{x},\tau) + 2(\psi(\vec{x},\tau)\,\delta_{ij} - \partial_i \partial_j E(\vec{x},\tau))
			\end{matrix} \right) 
\end{eqnarray}
The $\phi_{ij}$ are tensor modes (with respect to a three-dimensional spatial hypersurface) which
are traceless and transverse:
\begin{eqnarray}
	\delta_{ij} \, \phi_{ij} = 0 & \,\,\,& \partial_{i} \phi_{ij} = 0
\end{eqnarray}
The remaining 5 fields $\Phi,\psi,B,E,\chi$ are scalars. Two of them have familiar 
interpretations: the function $\Phi$, in the Newtonian limit, is the gravitational potential.
The function $\psi$ is known as the ``curvature perturbation" and is related to the intrinsic 
curvature $\tilde{R}$ on a constant $\tau$ hypersurface 
by:
\begin{eqnarray}\label{intrinsiccurvaturevar}
	\delta \tilde{R} & = & - \frac{4}{\,a(\tau)^2} \, \vec{\partial}^{\,2} \psi
\end{eqnarray}

Diffeomorphism invariance implies that the physical content of the fluctuations is limited to 
excitations which cannot be absorbed in a local change of coordinates. The tensor modes $\phi_{ij}$, which are invariant under such a change of coordinates, contain two physical degrees of freedom corresponding to the two independent polarizations of gravitational waves.
In addition there is a single physical scalar degree
of freedom which is represented by the five scalars $\Phi,\psi,B,E,\chi$.
Two of these fields can be eliminated by gauge conditions on the
coordinates, and two more will be removed by constraints, leaving one physical 
scalar. 

Unlike the tensor modes, the five scalars transform under a local change of coordinates. 
We are interested in identifying the gauge-invariant combinations of these fields that are related to 
measurable quantities. We can find the transformation properties of the scalars by writing out the transformation 
\eqref{mettr} of the metric, using the parametrization 
\eqref{scalarfluctuations}, and the scalar field under an infinitesimal diffeomorphism 
$x^{\mu} \rightarrow x^{\mu} + \eps^{\mu}$. This leads to the following transformation rules for the scalar perturbations:
\begin{eqnarray} \label{phitransform}
	\Phi &\rightarrow & \Phi + \HH \de \tau + (\de\tau)^\prime\\ \label{psitransform}
	\psi &\rightarrow & \psi - \HH \de \tau \\ \label{Btransform}
	B & \rightarrow & B - \ve^\prime + \de \tau \\ \label{Etransform}
	E & \rightarrow & E + \ve \\ \label{chitransform}
	\chi & \rightarrow & \chi -\vf^\prime \,\de \tau
\end{eqnarray}
where $\eps^{\tau} = \delta \tau$ and $\eps^{i} = \partial_{i} \ve$. It is straightforward 
to use these transformation rules
to identify gauge-invariant combinations of the scalar fluctuations. We will illustrate this by considering
four common gauge choices, each with a clear physical interpretation. The coordinate transformations required
to implement these gauge choices will help to identify a number of gauge-invariant quantities.

\subsubsection{Comoving Gauge}
Comoving gauge (or total matter gauge) 
is defined by the condition that an observer sees no local flux of energy-momentum due to the fluctuations in the fields. In other words, it is defined by $T_{\tau i} = 0$. For the spatially homogenous background the zeroth order part of $T_{\tau i}$ vanishes. The first order fluctuation in $T_{\tau i}$ is given by:
\begin{eqnarray} 
	\de T_{\tau i} & = & \vf' \, \partial_i \chi - a(\tau)^2 \, \partial_i B \, \left( -\frac{1}{2} \, \left(\frac{\vf'}{a}\right)^2 + V(\vf)\right)
\end{eqnarray}
In comoving gauge this should vanish. Starting in an arbitrary gauge we can impose the condition $\delta T_{i\tau}=0$ by making a coordinate transformation with:
 \begin{eqnarray} 
	\delta \tau = \frac{\chi}{\vf'} & & \partial_{\tau} \eps = B + \frac{\chi}{\vf'}
\end{eqnarray}
so that:
\begin{eqnarray} 
	\chi_{\rm com} & = &\chi - \vf' \de \tau \,\, = \,\, 0 \\
       B_{\rm com} & = & B - \partial_{\tau} \ve + \de \tau \,\, = \,\, 0
\end{eqnarray}
In this gauge $\psi$ becomes
\begin{eqnarray} \label{psicom}
	\psi_{\rm com} & = & \psi - \frac{\HH}{\vf'} \, \chi
\end{eqnarray}
Using the transformation rules \eqref{psitransform} and \eqref{chitransform} it is easy to verify that the variable $\psi_{com}$, the `comoving curvature perturbation', is itself gauge invariant. Comoving gauge can be thought of as the coordinate choice in which the curvature perturbation $\psi$ is equal to the gauge invariant quantity $\psi_{\rm com}$. 

\subsubsection{Flat Gauge}
\label{subsec:flatgauge}
Flat gauge is defined by $\psi_{\rm flat}=E_{\rm flat}=0$. In this gauge  
the perturbation $\de \tilde{R}$ of the spatial curvature vanishes\footnote{The condition \eqref{intrinsiccurvaturevar} as presented here only requires that $\psi=0$ for $\delta \tilde{R}$ to vanish. More generally, when the background involves a non-zero spatial curvature, the variation in the intrinsic curvature of a constant-time hypersurface will also depend on $E$.}. 
To reach flat gauge from a generic coordinate system transform $\psi$ and $E$ as:
\begin{eqnarray} 
	\psi_{\rm flat} & = & \psi - \HH \de \tau \,\, = \, \,0 \\
	E_{\rm flat} & = & E  + \ve \,\,=\,\,0
\end{eqnarray}
{\it i.e.} take $\delta \tau = \psi / \HH$ and $\ve = - E$. 
In flat gauge the observer does see a local flux in energy-momentum. It is due to 
a non-vanishing fluctuation in the scalar field, given by:
\begin{eqnarray} 
	\chi_{\rm flat} & = & \chi - \frac{\vf'}{\HH} \, \psi 
\end{eqnarray}
This variable is gauge invariant and is related to the comoving curvature perturbation via the relation:
\begin{eqnarray} 
	\chi_{\rm flat} & = & - \frac{\vf'}{\HH} \, \psi_{\rm com}
\end{eqnarray}
Except for a factor of $a(\tau)$ this is the same as 
Mukhanov's variable $\upsilon$ \cite{Mukhanov:jd}, which is given by:
\begin{eqnarray} 
	\upsilon & = & a\,\chi_{\rm flat}  =  - a\, \frac{\vf'}{\HH} \, \psi_{\rm com}= - 
	a \,\frac{\vf'}{\HH} \, \left( \psi - \frac{\HH}{\vf'}\, \chi\right)
\end{eqnarray}
Mukhanov's variable is often used to parameterize the scalar degree of freedom in inflation because its equation of motion is especially simple.

\subsubsection{Longitudinal Gauge}
The longitudinal (or conformal Newtonian) gauge is defined by a diagonal metric,
{\it i.e.} the off-diagonal components $E$ and $B$ vanish in longitudinal gauge.
Starting from an arbitrary gauge this condition determines the gauge parameters
as $\epsilon = -E$ and $\delta\tau=-(B+E^\prime)$. The non-trivial potentials
become:
\begin{eqnarray} 
	\Phi_{B} & = & \Phi - \HH (B+E') - B' - E'' \\
	\Psi_{B} & = & \psi + \HH (B+E')\\ 
	\chi_{B} & = & \chi + \vf' (B+E')
\end{eqnarray}
The metric functions $\Phi_{B}$ and $\Psi_{B}$ in longitudinal gauge are
the famous Bardeen variables \cite{Bardeen:kt}. 

\subsubsection{Uniform Energy Density Gauge}
Another physically interesting gauge is the gauge in which an oberserver sees no local 
variation in the energy density $\rho$. Since $\rho$ is a scalar the perturbation 
$\de \rho$ transforms as:
\begin{eqnarray} 
	\de \rho & \rightarrow & \de \rho - \rho' \, \de \tau
\end{eqnarray}
so $\de \rho_{\rm uni} = 0$ is reached by choosing $\de \tau = \de \rho /\rho^\prime$.
The curvature perturbation transforms as:
\begin{eqnarray}
	\psi_{\rm uni} & = & \psi - \frac{\HH}{\rho'}\, \de \rho
\end{eqnarray}
As before, this combination of variables is invariant under a gauge transformation, and we
can think of the Uniform Energy Density gauge as the coordinate system in which the
curvature perturbation $\psi$ is equal to the gauge-invariant quantity $\psi_{\rm uni}$.

We will not need $\psi_{\rm uni}$ in this paper and include it here only for completeness. However, we should point out that on super-horizon scales $\psi_{\rm uni}$ is essentially equal to $\psi_{\rm com}$. By computing $\de \rho$ we can compare $\psi_{\rm uni}$ and $\psi_{\rm com}$. The energy density appears on the right hand side of the Friedmann equation \eqref{metricEOM}. Its derivative is given by:
\begin{eqnarray}
    \rho\,' & = & -3 \, \HH \, \left(\frac{\vf\,'}{a}\right)^2
\end{eqnarray}
The variation in the energy density is given by:
\begin{eqnarray} 
	\de \rho & = & -3 \frac{\HH}{a^2} \, \vf' \, \chi - \frac{1}{a^2} \, \vec{\partial}^{\,2} \psi
\end{eqnarray}
Using these expressions and the definitions of 
$\psi_{\rm uni}$ and $\psi_{\rm com}$ gives:
\begin{eqnarray} 
	\psi_{\rm uni} & = & \psi_{\rm com} - \frac{1}{3 (\vf')^2} \, \vec{\partial}^{\,2} \psi
\end{eqnarray}
On super-horizon scales, where $\vec{k}^{\,2}$ is small compared to the Hubble scale, these variables are almost identical. 

\subsection{The Quadratic Action in Gauge-Invariant Variables}
\label{subsec:bndyaction}
We now discuss the on-shell action to quadratic order in the fluctuations 
around the background. The total action, including the contributions from the counterterms,
is given by:
\begin{eqnarray}
S_{\rm tot} & = & \int_{\MM_0} \bns d^4 x \sqrt{g}\,\left[ \frac{1}{2}\,R - \frac{1}{2} 
\, \nabla^{\mu} \vf
		\, \nabla_{\mu} \vf - V(\vf)\right] - 
		\int_{\dM_0} \bns d^3x \sqrt{\tg} \, K \\ \nonumber
		&   & - \int_{\dM_0} \bns d^3 x \sqrt{\tg} \, \left[ U(\vf) + M(\vf) \, \vec{D} \vf \cdot \vec{D} \vf
		 + C(\vf) \, \tilde{R} \right]
\end{eqnarray}
The first order variation of the action is:
\begin{eqnarray}
	\delta S_{\rm tot} & = & \int_{\MM_0} \bns d^4 x \sqrt{g} \left[ \frac{1}{2} \, h^{\mu\nu} \left( T_{\mu\nu}
	- G_{\mu\nu}\right) + \chi \, \left( \nabla^2 \vf - \pvf V\right)\right] \\ \nonumber
			&  & + \int_{\dM_0} \bns d^3 x \sqrt{\tg} \left[ h_{\mu\nu} \left( \pi^{\mu\nu}
				-P^{\mu\nu}\right) + \chi \, \left( \pi_{\vf} 
				- P_{\vf} \right)\right]
\end{eqnarray}
The on-shell quadratic action is then obtained from the variation of the first order terms, 
neglecting terms that vanish because they are proportional to the background equations of motion. This gives:
\begin{eqnarray} \label{secondvariation}
	\delta^{2} S_{\rm tot} & = & \int_{\MM_0} \bns d^4 x \sqrt{g} \, \left[ \frac{1}{2} \, h^{\mu\nu} \, \delta
	 \left(T_{\mu\nu} -G_{\mu\nu}\right) + \chi \, \delta \left( \nabla^2 \vf - \pvf V\right)\right] \\ \nonumber
		& & + \int_{\dM_0} \bns d^3 x \sqrt{\tg} \, \left[ h_{\mu\nu} \, \delta \left(  \pi^{\mu\nu}
				-  P^{\mu\nu}\right) + \chi \, \delta \left(
				\pi_{\vf} - P_{\vf} \right)\right]
\end{eqnarray}
Assuming a particular gauge before evaluating the variations in this expression 
would considerably simplify the calculation. However, choosing a gauge \emph{assumes} 
gauge invariance, which we would like to understand explicitly. We therefore evaluate 
the action without choosing a gauge:
\begin{eqnarray}\label{quadactionnogauge}
 \delta^{2} S_{\rm tot} & = & \int_{\MM_0} \bns d^4 x \sqrt{g} \, \frac{1}{a^2} \, \left[
 	 2 \,\Phi \,\vec{\partial}^{\,2} (\psi+\HH (B+E\,')) -4 \,\HH\,\Phi\,\left(\psi\,'+ \HH\,\Phi +
	\frac{1}{2} \,\vf\,'\,\chi\right)  \right. \\ \nonumber
  & & \hspace{20pt} - 2\,\HH\,\Phi\,\psi\,'-2\,\HH\,'\,\Phi^{2} + \vf\,'\,\Phi\,\chi\,' - \vf\,''\,\Phi\,\chi  -2 \, \partial_i B \, \partial_i \left( \psi\,'+\HH\,\Phi+\frac{1}{2}\,\vf\,'\,\chi\right) \\ \nonumber
  & & \hspace{20pt} + 2\,(3\psi-\vec{\partial}^2E) \, \left( \partial_\tau + 2 \HH \right) \left( \psi\,' + 
  	\HH\,\Phi + \frac{1}{2}\,\vf\,'\,\chi\right) \\ \nonumber
  & & \hspace{20pt} - 2\psi \, \vec{\partial}^{\,2} \left( \psi-\Phi+B\,'+E\,''+2\,\HH\,\left(B+E\,'\right)
  	\raisebox{15pt}{\,} \right)	\\ \nonumber
    & & \hspace{20pt} - \chi \, \chi\,'' - 2\,\HH\,\chi\,\chi \,' - a^2 \, \pvf^{\,2}V \, 
  \chi^2  + \chi\,\vec{\partial}^{\,2}\left( \chi + \vf\,' \, (B+E\,')\right)\\ \nonumber	
  & & \hspace{20pt}  -2\,(\vf\,''+2\,\HH\,\vf\,')\,\Phi\,\chi
  -\vf\,'\,\chi\,\Phi\,'-3\,\vf\,'\,\chi\,\psi\,' \\ \nonumber
  & & \hspace{15pt} \left. - \frac{1}{4} \, \phi_{ij} \, \left( \phi_{ij}\,'' + 2\,\HH\,\phi_{ij}\,'
       -\vec{\partial}^{\,2} \phi_{ij} \right) \raisebox{15pt}{\,}\right]
\end{eqnarray}
\begin{eqnarray}   \nonumber
  & & + \int_{\dM_0} \bns d^3 x \sqrt{\tg}\, \left[ \raisebox{15pt}{\,}
  	\frac{1}{a}\, \chi \,\chi\,' + \frac{\vf\,'}{a} \,\chi\,\Phi - 
  	\pvf^{\,2} U \, \chi^2 + 2\,M\,\chi \,\vec{D}^{\,2} \chi \right. \\ \nonumber 
  & & \hspace{20pt} + 4\,\pvf C\,\left( \chi \, \vec{D}^{\,2} \psi + \psi\,\vec{D}^{\,2}\chi \right) + \left( 
  	3\,\frac{\vf\,'}{a} - 6\,\pvf U \right)\,\chi\,\psi \\ \nonumber
  & & \hspace{20pt} + \left( 2 \, \pvf U - \frac{\vf'}{a} \right) \, \chi\,\vec{\partial}^{\,2} E - 
  	\frac{6}{a} \, \psi\,\psi\,' - 3\,\left( U + 2 \,\frac{\HH}{a} \right)\,\psi^{2} \\ \nonumber
  & & \hspace{20pt} - 6\,\left( U+2\, \frac{\HH}{a} \right) \, \psi\,\vec{\partial}^{\,2} E 
      + 4\,C\,\psi\,\vec{D}^{\,2} \psi -6\,\frac{\HH}{a}\,\psi\,\Phi\\ 
      \nonumber
  & & \hspace{20pt}  + 2 \,\frac{\HH}{a}\,\Phi\,\vec{\partial}^{\,2} E + 
  	\frac{2}{a} \, \psi\,\vec{\partial}^{\,2} B + \left( U + 2\frac{\HH}{a}\right) \,
	\vec{\partial}^{\,2} E \,\vec{\partial}^{\,2} E \\ \nonumber
  & & \hspace{20pt} + \frac{2}{a}\,\left( \psi\,'\,\vec{\partial}^{\,2}E + \psi\,
  	\vec{\partial}^{\,2} E\,'\right) + \frac{1}{4\,a} \, \phi_{ij}
	\,\phi_{ij}\,'-\frac{1}{2}\,\left(U+2\,\frac{\HH}{a}\right)\,\phi_{ij}\,\phi_{ij}
	\\ \nonumber
  & & \hspace{20pt} \left. - \frac{1}{2\,a^2} \, C\,\phi_{ij} \, \vec{\partial}^{\,2} \phi_{ij}
   \raisebox{15pt}{\,} \right]
\end{eqnarray}
Terms proportional to the background equations of motion have been cancelled in this 
expression, but all other terms have been left intact. Specifically, several terms involving 
$U$ and $\pvf U$ can be see to cancel after substituting the values 
$U=-2 \HH / a$ and $\pvf U = \vf\,' / a$. Those terms are left explicit here so that we easily isolate the contributions of the counterterms to the quadratic action.
By setting $U$, $M$, and $C$ to zero we remove all such contributions, and are left with only those terms that came from the original action \eqref{inflationaction}.

The focus of the next section will be explicitly demonstrating that this action is only gauge-invariant when we include the contributions from the counterterms. But let us \emph{momentarily} assume gauge invariance so that we may immediately present the action \eqref{quadactionnogauge} in gauge invariant variables, suitable for application to problems in inflation and cosmology. Since the tensor modes are gauge invariant already we can simply note their contribution to \eqref{quadactionnogauge}:
\begin{eqnarray}\label{tensorAction} 
 S[\phi_{ij}] & = & - \int_{\MM_0} \bns d^{4}x \sqrt{g} \, 
 \frac{1}{8\,a^2} \, \eta^{\mu\nu} \, \partial_{\mu} \phi_{ij} 
 \, \partial_{\nu} \phi_{ij}
 + \int_{\dM_0} \bns d^{3}x \sqrt{\tg} \, \frac{C}{4\,a^2} 
 	 \, \vec{\partial} \phi_{ij} \cdot \vec{\partial} \phi_{ij} 
\end{eqnarray}
To obtain this expression we have integrated the tensor terms in \eqref{quadactionnogauge} by parts. Although the tensor modes carry indices, we write their kinetic term using partial derivatives to emphasize the fact that the two polarization states of the tensor modes are
equivalent to two minimally coupled, massless scalars.


It is straightforward to write the action for the scalar fluctuations in terms of gauge-invariant variables. Starting with \eqref{quadactionnogauge} the first step is to eliminate two of the scalars $\Phi, \psi, B, E, \chi$ by making a gauge choice. The variables remaining after gauge fixing are related by two constraints, which can be succinctly expressed in terms of gauge invariant variables as:
\begin{eqnarray}
\label{gaugeconstrBardeen}
\Psi_B^\prime  + {\cal H}\,\Phi_B + \frac{1}{2}\,\varphi^\prime\,\chi_B & = & 0 \\ \nonumber
\Phi_B -\Psi_B & = & 0
\end{eqnarray} 
In an arbitrary gauge these equations take the form: 
\begin{eqnarray}
\label{gaugeconstr}
\psi^\prime  + {\cal H}\Phi + \frac{1}{2}\varphi^\prime\chi & = & 0 \\ \nonumber
\Phi - {\cal H}(B +E^\prime) - (B+E^\prime)^\prime & = & \psi + {\cal H}(B +E^\prime)
\end{eqnarray} 
After these constraints are imposed only one physical degree of freedom will 
remain. 

For example, in comoving gauge, take $\chi_{\rm com}=B_{\rm com}=0$ and 
then use the constraints \eqref{gaugeconstr} to write the remaining expression 
in terms of the comoving curvature perturbation $\psi_{\rm com}$. The result is:
\begin{eqnarray}\label{psicomAction}
 S[\psi_{\rm com}] & = & - \int_{\MM_0} \bns d^4 x \sqrt{g} \,\, \frac{1}{2} 
   \left( \frac{\vf\,'}{\HH}\right)^2 \, \nabla^{\mu} \psi_{\rm com} \, \nabla_{\mu}
   \psi_{\rm com} 
   \\ \nonumber & & 
 - \int_{\dM_0} \bns d^3 x \sqrt{\tg} \,\frac{1}{a^2}\, \left(\frac{\vf\,'}{\HH}\right)^2 \frac{U}{\pvf U}
   \, \pvf C \, \vec{\partial} \psi_{\rm com} \cdot \vec{\partial} \, \psi_{\rm com}  
\end{eqnarray}
We can also perform this computation in flat gauge. In this case the action, written in 
terms of Mukhanov's variable $\upsilon$, becomes:
\begin{eqnarray}\label{MukhanovAction}
  S[\ups] & = &  \int_{\MM_0} \bns d^4 x \left[ -\frac{1}{2}\,\eta^{\mu\nu} \partial_{\mu} \ups 
          \partial_{\nu} \ups + \frac{1}{2}\,\frac{z''}{z}\,\ups^2\right]
	 + \int_{\dM_0} \bns d^{3} x \left[ 
	  -\frac{1}{2}\,\frac{z\,'}{z}\, \ups^2 - \frac{1}{a} \, M \vec{\partial} \ups \cdot \vec{\partial}\ups\right]
\end{eqnarray}
where $z$ is defined as $z = a \,\frac{\vf\,'}{\HH\,}$. The expressions \eqref{tensorAction},
\eqref{psicomAction}, and \eqref{MukhanovAction} for the quadratic action in the 
presence of a boundary are new results, as far as we are aware. 

In our previous paper \cite{Larsen:2003pf} we computed the quadratic action in longitudinal gauge. The action was then put on-shell by explicitly solving the bulk equations of motion
and evaluating the boundary terms on the bulk solution. The equations of motion are difficult to solve in general, which limited our calculation to the case of slow-roll inflation. 
In the present paper we will instead use symmetries to determine the scaling behavior of the 
on-shell action, allowing us to partially circumvent the problem of solving the bulk equations explicitly. 

\subsection{Gauge Invariance of the Quadratic Action}
In the preceding subsection we computed the quadratic action in specific gauges, 
with gauge invariance temporarily assumed. However, as we have emphasized 
repeatedly, diffeomorphism invariance is not automatic in the presence of a boundary.
Indeed, if we neglect counterterms then the quadratic action 
\eqref{quadactionnogauge} is {\it not} gauge invariant. The simplest way to demonstrate this is to consider 
fluctuations which are pure gauge:
\begin{eqnarray} \label{gaugemodes}
	\Phi & = & \HH \, \delta\tau + (\delta\tau)^\prime \\ \nonumber
	\psi & = & - \HH \,\delta \tau \\ \nonumber
	B & = & - \ve^\prime + \delta \tau \\ \nonumber
	E & = & \ve \\ \nonumber
	\chi & = & - \vf\,' \, \delta \tau
\end{eqnarray}
Inserting these modes in  \eqref{quadactionnogauge} gives:
\begin{eqnarray}
	\delta_{\eps}^{\,2} S & = & \int_{\dM_0} \bns d^3 x\sqrt{\tg} \,\frac{1}{a}\,\left[
	(4\,\HH^3 - 8\,\HH\,\HH\,'-\HH\,'')\delta\tau^2-2{\cal H}\,\delta\tau\,\vec{\partial}^{\,2}\delta\tau+\right.
	\\
	&~&\left. +2\,{\cal H}\,\vec{\partial}^{\,2}\epsilon\,\vec{\partial}^{\,2}\epsilon
	+\left( 16\,\HH^2 - 4\,\HH\,'\right) \delta\tau\,\vec{\partial}^{\,2}\epsilon
	\right]\nonumber
\end{eqnarray}
This expression clearly does not vanish in anything but exceptional circumstances. The 
failure of the quadratic action to vanish for pure gauge modes exhibits a violation of diffeomorphism invariance.  

The introduction of counterterms can rectify this problem. Inserting the gauge modes \eqref{gaugemodes} in the quadratic action \eqref{quadactionnogauge} we 
find that almost all of the terms cancel due to the background equations of motion 
and the expression 
\eqref{Uctdiff} for the counterterm $U(\varphi)$. The result is now:
\begin{eqnarray}
	\delta_{\eps}^{\,2} S_{\rm tot} & = & - \int_{\dM_0} \bns d^3 x \sqrt{\tg} \,2\,\frac{\HH}{a} \, \left( 1 + U\,C 
	-4\,\pvf U\, \pvf C + 2 M\,\frac{(\pvf U)^2}{U}\right)\,\delta \tau \, \vec{\partial}^{\,2}\,\delta \tau
\end{eqnarray}
Thus, the total action $S_{\rm tot}$ is gauge invariant if the counterterms satisfy the equation:
\begin{eqnarray}
\label{UMCeqn}
   1 + U\,C -4\,\pvf U\, \pvf C + 2 M\,\frac{(\pvf U)^2}{U} & = & 0
\end{eqnarray}
This condition must be satisfied by the counterterms in order that the full quadratic
action vanishes for pure gauge modes; {\it i.e.} it is a necessary condition for
diffeomorphism invariance. We will show below that this condition is satisfied
by the counterterms \eqref{Ceqn} and \eqref{Meqn} that we found by cancelling 
divergences.

In the preceding subsection we presented the quadratic action in two forms:
\eqref{psicomAction} was written in terms of the comoving curvature 
perturbation $\psi_{\rm com}$, and \eqref{MukhanovAction} was written in terms of 
Mukhanov's variable $\upsilon$. These actions should, of course, be equivalent. 
We can try to verify this by noting that $\psi_{\rm com}$ and $\upsilon$ are related 
as:
\begin{eqnarray}
\label{upsilonpsirelation}
	\upsilon & = & -a\,\frac{\vf\,'}{\HH} \, \psi_{\rm com}
\end{eqnarray}
But this simple substitution is {\it not} enough to relate the actions. Indeed,
the action \eqref{psicomAction} depends on the counterterms $U(\vf)$ and 
$C(\vf){\cal R}$, but not $M(\vf)\vec{D}\varphi\cdot\vec{D}\varphi$. This is because
in comoving gauge the spatial inhomogeneties are contained entirely in the fluctuations
of the metric, so that the counterterm involving $M(\vf)$ does not contribute to the action.
The action \eqref{MukhanovAction}, on the other hand, depends on the counterterms $U(\vf)$ and 
$M(\vf)\vec{D}\varphi\cdot\vec{D}\varphi$, but not $C(\vf){\cal R}$. In flat gauge the
spatial inhomogeneities are parameterized by $\chi_{\rm flat}$ and the action does not depend on the $C(\vf)$ counterterm. Therefore, the two actions can agree only if $C(\vf)$ and $M(\vf)$ are related. More precisely, if we have:
\begin{eqnarray} 
\label{MCrelation}
	M(\vf) & = & \frac{U(\vf)}{\pvf U(\vf)} \, \pvf C(\vf)
\end{eqnarray}
then the two actions are in fact identical upon the substitution \eqref{upsilonpsirelation}. 

The relations \eqref{UMCeqn} and \eqref{MCrelation} are conditions that the counterterms
must satisfy in order for the total action to be diffeomorphism invariant. We can simplify the 
condition \eqref{UMCeqn} using \eqref{MCrelation} and find:
\begin{eqnarray}
	1 + U\,C - 2\,\pvf U \, \pvf C & = & 0
\end{eqnarray}
which is precisely $\eqref{Ceqn}$. Furthermore, we can rearrange:
\begin{eqnarray}
   \left( 1 - \frac{U}{\pvf U} \, \pvf \, \right) \, \left( 1 + U \, C - 2\,\pvf U \, \pvf C \right) & = & 0
\end{eqnarray}
using \eqref{MCrelation} and then recover $\eqref{Meqn}$. \emph{Diffeomorphism invariance
of the total action guarantees that the counterterms satisfy the equations 
$\eqref{Ueqn}$, $\eqref{Ceqn}$, $\eqref{Meqn}$, which were originally motivated
by the cancellation of divergences}. 

One might wonder if these conditions on the counterms, necessary for diffeomorphism invariance,
are in fact \emph{sufficient} to guarantee a diffeomorphism invariant action. The most straightforward way to show that this is indeed the case is to use the expressions \eqref{UMCeqn}
and \eqref{MCrelation} in the action \eqref{quadactionnogauge}, where the gauge has not been fixed. Though tedious, after simplifying the action using the constraints \eqref{gaugeconstrBardeen} the various terms combine to form gauge invariant variables. As a check of our calculations, the actions in comoving gauge \eqref{psicomAction} and flat gauge \eqref{MukhanovAction} were both obtained in this manner.

Let us summarize the main results of our discussions of counterterms at different 
points in sections \ref{bndyctterms} and \ref{sec:DiffInv}. There are (at least) two ways 
to motivate the addition of local counterterms on the boundary, and determine their form:
\begin{enumerate}
\item
The on-shell action diverges as $\tau_0\to 0$. The counterterms are needed
to cancel these divergences. The H-J equation shows that the functions 
$U(\varphi),M(\varphi),C(\varphi)$ defining the counterterms must satisfy 
eqs. $\eqref{Ueqn}$, $\eqref{Ceqn}$, $\eqref{Meqn}$.

\item
Diffeomorphism invariance cannot be taken for granted in the presence of 
a boundary. Counterterms are needed to restore diffeomorphism invariance. 
We implement diffeomorphism invariance by imposing cancellation of 
boundary sources, vanishing action for fluctuations of pure gauge form, and
equivalence of boundary actions in comoving and flat gauges; these
imply that the counterterms must satisfy eqs. \eqref{Uctdiff}, \eqref{UMCeqn}, 
and \eqref{MCrelation}. 
\end{enumerate} 
The three conditions imposed on counterterms under (1) are equivalent to
those imposed under (2). We can thus take either (1) or (2) as the 
primary motivation, the other will then follow. More precisely, starting 
with either (1) or (2), the other is in fact a consistency condition: the H-J equation 
is a manifestation of the Hamiltonian constraint, so diffeomorphism invariance 
is assumed, albeit not emphasized, in (1). Similarly, once we have imposed 
diffeomorphism invariance in (2), the Hamiltonian constraint is satisfied and 
so divergences are cancelled. The real lesson is that the counterterms 
are an essential ingredient for a consistent gravitational theory.

\section{The Holographic Renormalization Group}
\label{sec:consequences}
In this section we exploit diffeomorphism invariance to constrain the functional 
form of the on-shell action and its functional derivatives. 
An especially in important class of four dimensional diffeomorphisms act as 
Weyl rescalings of the boundary data. These ensure that the on-shell action satisfies a renormalization-group equation and that second order terms in the action, describing 
fluctuations around the background, obey a Callan-Symanzik equation. 
These constraints on the action suggest an analogy 
between inflation and a three-dimensional Euclidean field theory near a renormalization 
group fixed point, with the on-shell action interpreted as the generating functional for 
correlators of field theory operators sourced by the boundary data.

\subsection{The Master Equation}

The on-shell, renormalized action is a functional of the induced 
metric\footnote{In the remainder of the paper we will allow a slightly different notation for the induced 
metric and other tensors intrinsic to the $\tau=\tau_0$ hypersurface. In order to highlight the difference 
between the directions normal to and parallel to the boundary, we will use latin indices $i,j,\ldots$ on 
intrinsic tensors. This should \emph{not} be interpreted as four dimensional tensors simply restricted 
to their spatial components.} 
$\tg_{ij}(\vec{x},\tau)$ and the field $\vf(\vec{x},\tau)$
evaluated on the spacelike hypersurface $\tau=\tau_0$:
\begin{eqnarray}
	S_{\rm tot} & = & S_{\rm tot} \left[ \vf(\vec{x},\tau_0), \tg_{ij}(\vec{x},\tau_0) \right]
\end{eqnarray} 
In the following sections we will use the notation $\vf(\vec{x})$ to refer to $\vf(\vec{x},\tau_0)$, and similarly for $\tg_{ij}(\vec{x})$.

Diffeomorphism invariance requires that the transformation of $S_{\rm tot}$ under 
an infinitesimal four dimensional diffeomorphism vanishes:
\begin{eqnarray}
	\delta_{\eps} S_{\rm tot} & = & \int d^3 x \, \left[ \delta_{\eps} \tg_{ij} \, 
	\frac{\delta S_{\rm tot}}{\delta \tg_{ij}} + \delta_{\eps} \vf \, 
	\frac{\delta S_{\rm tot}}{\delta \vf} \right] =0 
\end{eqnarray}
The transformation laws of the fields \eqref{mettr} and \eqref{scalartr} can be written 
in terms of transverse and normal components as:
\begin{eqnarray}
	\delta_{\eps} \vf & = & n_{\lambda} \eps^{\lambda} \, n^{\mu} \nabla_{\mu} \vf 
		- \eps^i D_i \vf \\ \nonumber
	\delta_{\eps} \tg_{ij} & = & - D_i \eps_j - D_j \eps_i - 2 \, n_{\lambda} \eps^{\lambda} K_{ij}
\end{eqnarray}
The first two terms in the transformation of $\tg_{ij}$ are due to the spacelike part of the diffeomorphism $\eps^{i}$. The last term appears because the extrinsic curvature is 
essentially the normal derivative of the induced metric on a hypersurface. 
The transformation of the on-shell action becomes:
\begin{eqnarray}
  \delta_{\eps} S_{\rm tot} & = & \int d^3 x \left[ -2 D_i \eps_j (\vec{x}) \, 
    \frac{\delta S_{\rm tot}}{\delta \tg_{ij}(\vec{x})} - \eps^i (\vec{x}) \, 
    D_i \vf(\vec{x}) \, \frac{\delta S_{\rm tot}}{\delta \vf(\vec{x})} \right.
    \\ \nonumber
  & & \hspace{20pt} \left. + n_{\lambda} \eps^{\lambda}(\vec{x}) \left( n^{\mu} 
    \nabla_{\mu} \vf(\vec{x}) \, \frac{\delta S_{\rm tot}}{\delta \vf(\vec{x})}
    -2\,K_{ij} \, \frac{\delta S_{\rm tot}}{\delta \tg_{ij}(\vec{x})} \right)\right]
   =  0 \nonumber
\end{eqnarray}
The field transformations can be written as:
\begin{eqnarray}
  \delta_{\eps} \vf(\vec{x}) & = & - \eps^i \, D_i \vf(\vec{x}) 
  - \lambda_{\eps} \, \beta(\vf(\vec{x}))\\
\delta_{\eps} \tg_{ij} & = &  -  D_i \eps_j - D_j \eps_i - 2 \, \lambda_{\eps} \, \tg_{ij}
\end{eqnarray}
where $\lambda_{\eps} = \HH \, \delta \tau$ parametrizes the variation of the normal,
and we defined $\beta(\vf)$ as:
\begin{eqnarray}\label{betadefn}
   \beta(\vf) & = & \frac{1}{\HH} \, \frac{\partial \vf}{\partial \tau}
\end{eqnarray}
Expressed this way, the transformation properties of $\vf(\vec{x})$ and $\tg_{ij}(\vec{x})$ under a four dimensional diffeomorphism have a clear interpretation on the $\tau=\tau_0$ hypersurface: diffeomorphisms involving the direction normal to the hypersurface are realized as Weyl rescalings, while diffeomorphisms involving directions along the hypersurface are 
interpreted as three-dimensional diffeomorphisms. The transformation of the on-shell action 
under diffeomorphisms can now be written:
\begin{eqnarray}\label{mastereqn}
  \delta_{\eps} S_{\rm tot} & = & \int d^3 x \left[ -2 D_i \eps_j (\vec{x}) \, 
    \frac{\delta S_{\rm tot}}{\delta \tg_{ij}(\vec{x})} - \eps^i (\vec{x}) \, 
    D_i \vf(\vec{x}) \, \frac{\delta S_{\rm tot}}{\delta \vf(\vec{x})} \right.
    \\ \nonumber
  & & \hspace{20pt} \left. - \lambda_{\eps}(\vec{x}) \left( \beta 
    (\vf(\vec{x})) \, \frac{\delta S_{\rm tot}}{\delta \vf(\vec{x})}
    +2\,\tg_{ij}(\vec{x}) \, \frac{\delta S_{\rm tot}}{\delta \tg_{ij}(\vec{x})} 
    \right)\right] = 0 \nonumber
\end{eqnarray}
This is our master equation which constrains the dependence of $S_{\rm tot}$ on the 
fields $\vf(\vec{x})$ and $\tg_{ij}(\vec{x})$. We simplified the equation by writing the
extrinsic curvature as $K_{ij} = (\HH/a) \, \tg_{ij}$.

As it stands the master equation \eqref{mastereqn} is a functional differential equation 
for the action $S_{\rm tot}$. Its
meaning becomes more transparent when we exploit the 
gauge invariance of $S_{\rm tot}$ and choose a simple gauge. For example, 
in the flat gauge described in section \ref{subsec:flatgauge},
the spatial inhomogeneities due to fluctuations around the background are 
contained entirely in the scalar field\footnote{Here we have again used a notation 
that leaves the dependence on $\tau_0$ implicit, with $\vf$ representing the 
background value of the scalar.}:
\begin{eqnarray}
	\vf_{f}(\vec{x}) & = & \vf + \chi_{f}(\vec{x})
\end{eqnarray}
In this gauge the induced metric on $\dM_0$ is simply $\tg_{ij} = a(\tau_0)^2 \delta_{ij}$;
so, in flat gauge, the total action is a functional of the scalar $\vf_{f}(\vec{x})$ and a \emph{function} of the scale factor $a(\tau_0)$. We can expand the action as a series in $\chi_{f}(\vec{x})$:
\begin{eqnarray}\label{actionexpansion}
  S_{\rm tot}[a, \vf_{f}(\vec{x})] & = & S_{\rm tot}^{(0)} + \sum_{n=1}^{\infty}  \int \prod_{i=1}^{n}
  d^3 x_i \, \frac{1}{n!} \, S_{\rm tot}^{(n)}[a,\vf;\vec{x}_1,\ldots,\vec{x}_n] \, 
  \chi_{f}(\vec{x}_1) \ldots \chi_{f}(\vec{x}_n) 
\end{eqnarray}
where the coefficient $S_{\rm tot}^{(n)}$ is the $n^{th}$ functional derivative of $S_{\rm tot}$ 
with respect to $\vf_{f}(\vec{x})$, evaluated at 
$\phi_{f}(\vec{x}) = \phi$ (or $\chi_{f}(\vec{x}) = 0$):
\begin{eqnarray}
  S_{\rm tot}^{(n)}[a,\vf;\vec{x}_1,\ldots,\vec{x}_n] & = & 
  	\frac{\delta^{n}\, S_{\rm tot}}{\delta \vf_{f}(\vec{x}_1) \ldots \delta \vf_{f}(\vec{x}_n)}
	\left. \raisebox{15pt}{\,}\right\vert_{\chi_{f} = 0}
\end{eqnarray}
The first term in the expansion, $S_{\rm tot}^{(0)}$, is simply the action for the background; 
{\it i.e.} the part of the action with no dependence on the fluctuations. Furthermore, 
since the action is on-shell its first order variation must vanish for arbitrary fluctuations 
around the background, implying $S_{\rm tot}^{(1)} = 0$.

In the following we will write the master equation \eqref{mastereqn} in more
explicit forms for specific choices of the diffeomoprphism $\eps^{\mu}$. 
This will lead to differential equations that constrain the dependence of the coefficients
$S_{\rm tot}^{(n)}$ on the homogeneous background fields $a$, $\vf$, and the $n$ 
spatial points $\vec{x}_{i}$.

\subsubsection{Weyl Rescalings}
First we consider a diffeomorphism that only involves the normal direction and is 
independent of the spatial coordinates. Then $\vec{\eps} \, = \, 0$ and 
$\lambda_{\eps}$ depends only on $\tau_0$. 
Under this diffeomorphism the master equation \eqref{mastereqn} becomes:
\begin{eqnarray}\label{functionalrgeqn}
  \delta_{\eps} S_{\rm tot} & = & - \lambda_{\eps} \, \int d^3 z \left( 2 \, \tg_{ij}(\vec{z}) 
    \, \frac{\delta S_{\rm tot}}{\delta \tg_{ij}(\vec{z})} + \beta(\vf(\vec{z})) \, \frac{\delta 
    S_{\rm tot}}{\delta \vf(\vec{z})}\right) \,\,\, = \,\,\, 0
\end{eqnarray}
In flat gauge the action is a \emph{function} of the scale factor $a(\tau_0)$ and a 
functional of $\vf_{f}(\vec{x})$; so it is more appropriate to write \eqref{functionalrgeqn} 
as:
\begin{eqnarray}
  \left( a \, \frac{\partial}{\partial a} + \int d^3 z \, 
    \beta(\vf_{f}(\vec{z})) \, \frac{\delta }{\delta \vf_{f}(\vec{z})} \right) \, 
    	S_{\rm tot} & = & 0
\label{masterweyl}
\end{eqnarray}

We evaluate this equation for the background by setting the fluctuations equal to 0. In
this case the complete scalar field $\vf_f$ reduces to the homogeneous background
$\vf$ and we can treat $S_{\rm tot}$ as a function of $\vf$, rather than a functional.
We can therefore write \eqref{masterweyl} as:
\begin{eqnarray}\label{rgeqn}
  \left( a \, \frac{\partial}{\partial \, a} + \beta(\vf) \, \frac{\partial}{\partial \vf} \,\right) 
S_{\rm tot}^{(0)} & = & 0
\end{eqnarray}
Equation \eqref{rgeqn} can be interpreted as an RG equation for $S_{\rm tot}^{(0)}$. The
field $\vf$ plays the role of a coupling with beta function $\beta(\vf)$,
and $a(\tau_0)$ is the scale factor. The $\beta$-function was defined 
in \eqref{betadefn} as $\beta = {\cal H}^{-1}\partial_\tau\vf$ so the RG 
equation \eqref{rgeqn} states that the $\tau_0$-dependence of 
$S_{\rm tot}^{(0)}$ through the scale factor $a(\tau_0)$ is balanced by the dependence 
of $\tau_0$ through the background scalar field $\vf(\tau_0)$. We can express this 
scale independence as: 
\begin{equation}
\frac{dS_{\rm tot}^{(0)}}{da}=0
\end{equation}
or, alternatively, as independence of the cut-off $\tau_0$:
\begin{equation} 
\frac{dS_{\rm tot}^{(0)}}{d\tau_0}=0
\end{equation}

We want to find additional differential equations which constrain the higher
coefficients  $S_{\rm tot}^{(n)}$ in the expansion \eqref{actionexpansion} of the action.
To do so we must vary  \eqref{functionalrgeqn} with respect to the
field $\vf_f$, before taking $\chi_f=0$. Taking one functional derivative 
of \eqref{functionalrgeqn} with respect to $\vf_{f}(\vec{x})$ 
yields the functional equation:
\begin{eqnarray}\label{firstfunctionaleqn}
  W_{1}[\vf_{f}(\vec{x})] \cdot \frac{\delta S_{\rm tot}}{\delta \vf_{f}(\vec{x})} & = & 0
\end{eqnarray}
where $W_{1}[\vf_{f}(\vec{x})]$ has the form:
\begin{eqnarray}
  W_{1}[\vf_{f}(\vec{x})] & = & a \, \frac{\partial}{\partial \,a} + \int d^3 z \, 
    \left( \beta(\vf_{f}(\vec{z})) \, \frac{\delta }{\delta \vf_{f}(\vec{z})} 
    + \delta(\vec{z}-\vec{x}) \, \frac{\delta \beta(\vf_{f}(\vec{z}))}{\delta \vf_{f}(\vec{z})}  
    \right)
\end{eqnarray}
Evaluating \eqref{firstfunctionaleqn} for the background by setting $\chi_{f}(\vec{x}) = 0$,
we can replace the functional derivatives by ordinary derivatives, and
find the differential equation:
\begin{eqnarray}\label{trivialCSeqn}
   \left( a \, \frac{\partial}{\partial \, a} + \beta(\vf) \, \frac{\partial}{\partial \vf} \, + 
   \gamma(\vf)\right) S_{\rm tot}^{(1)}[a,\vf;\vec{x}] & = & 0
\end{eqnarray}
where we defined the anomalous dimension as:
\begin{equation}
\label{gammadef}
\gamma(\vf)\equiv \frac{\partial \beta(\vf)}{\partial \vf}
\end{equation}
Equation \eqref{trivialCSeqn} takes the form of a Callan-Symanzik equation 
in which the function $S_{\rm tot}^{(1)}$ plays the role of a one-point
function of an operator with anomalous dimension $\gamma$.
Of course this equation is actually trivial since, as discussed in the previous subsection, 
the expansion for $S_{\rm tot}$ does not contain a linear term and so 
$S_{\rm tot}^{(1)}=0$. 

However, we can repeat the procedure by taking one more 
functional derivative of \eqref{firstfunctionaleqn} with respect to $\vf_{f}(\vec{y})$ 
and evaluating the resulting functional differential equation at $\chi_{f} = 0$. 
This yields an equation satisfied by $S_{\rm tot}^{(2)}$:
\begin{eqnarray}\label{1stCSeqn}
  \left( a\,\frac{\partial}{\partial a} + \beta(\vf) \, \frac{\partial}{\partial 	    \vf} + 2 \gamma(\vf) \right) \, 
    S_{\rm tot}^{(2)}[a,\vf;\vec{x},\vec{y}] & = & 0
\end{eqnarray}
A contact term proportional to $\delta(\vec{x}-\vec{y})\partial_\vf^2\beta$ was
omitted because it acts on $S_{\rm tot}^{(1)}$ and so vanishes. 
Just as equation \eqref{trivialCSeqn} can be thought of as the Callan-Symanzik equation
for the one-point function of an operator with anomalous dimension $\gamma$, equation \eqref{1stCSeqn} can be regarded as the Callan-Symanzik equation for the two-point function of the same operator. In this case, $S_{\rm tot}^{(2)}$ corresponds to the two-point function. 

\subsubsection{Three Dimensional Diffeomorphisms}
A particularly simple case of the master equation \eqref{mastereqn} corresponds to three dimensional diffeomorphisms. These are generated by four dimensional diffeomorphisms with:
\begin{eqnarray}
   n_{\mu} \eps^{\mu} = 0
\end{eqnarray}
In this case \eqref{mastereqn} becomes:
\begin{eqnarray}
\delta_{\vec{\eps}} S_{\rm tot} & = & \int d^3 x \left[ -2 D_i \eps_j \,       
  \frac{\delta S_{\rm tot}}{\delta \tg_{ij}} - \eps^i \, D_i \vf \, \frac{\delta 
  S_{\rm tot}}{\delta \vf}\right] \, \, \, = \,\,\, 0
\end{eqnarray}
Integrating the first term by parts and requiring the integrand to vanish for arbitrary $\vec{\eps}$ leads to:
\begin{eqnarray}\label{3Ddiffeoeqn}
 2 \, D_j \left( \frac{\delta S_{\rm tot}}{\delta \tg_{ij}}\right) - D^i \vf \, 
 \frac{\delta S_{\rm tot}}{\delta \vf}  & = & 0
\end{eqnarray}
In the Hamilton-Jacobi formalism functional derivatives of the on-shell action with respect to the boundary data yield the canonical momenta, evaluated at the boundary. It is straightforward to see that \eqref{3Ddiffeoeqn} is simply the usual momentum constraint that appears in the canonical $3+1$ treatment of four dimensional gravity. Here it represents invariance of the total action under reparameterizations of the spatial coordinates.

\subsubsection{Conformal Transformations}
Weyl rescalings are diffeomorphisms that act only on the normal coordinate;
and three dimensional diffeomorphisms are bulk diffeomorphisms acting only
on the spatial coordinates. We can also consider four dimensional diffeomorphisms 
for which the induced three dimensional diffeomorphism compensates for a 
Weyl rescaling of the metric $\tg_{ij}$ in such a way that $\delta_{\eps} \tg_{ij} = 0$.
These diffeomorphisms act as conformal transformations on the boundary. 

Under a general diffeomorphism the induced metric on $\dM_{0}$ transforms as:
\begin{eqnarray}
  \delta_{\eps} \tg_{ij} & = & - D_i \eps_j - D_j \eps_i - 2 \, \lambda_{\eps} \tg_{ij}
  \,\,\, = \,\,\, 0
\end{eqnarray}
Therefore, the condition for $\delta_{\eps} \tg_{ij} = 0$ is:
\begin{eqnarray}
  D_i \eps_j + D_j \eps_i & = & -2\,\lambda_{\eps} \, \tg_{ij} \\ \nonumber
  D_i \eps^i & = & -3 \, \lambda_{\eps} 
\end{eqnarray}
For this type of diffeomorphisms the master equation \eqref{mastereqn} becomes:
\begin{eqnarray}
   \delta_{\eps} S_{\rm tot}= - T_{\eps}\left[ \vf \right] \cdot S_{\rm tot} & = & 0
\end{eqnarray}
where $T_{\eps} \left[ \vf \right]$ is the functional differential operator:
\begin{eqnarray}
   T_{\eps} \left[ \vf \right] & = & \int d^3 x \left[ \eps^i(\vec{x}) D_i \vf + 
  \lambda_{\eps} (\vec{x}) \, \beta(\vf) \right] \, \frac{\delta}{\delta \vf}
\end{eqnarray}
Applying \emph{two} functional derivatives, with respect to $\vf(\vec{x})$ 
and $\vf(\vec{y})$, we find:
\begin{eqnarray}
\label{tepsxy}
  T_{\eps}\left[ \vec{x},\vec{y},\vf \right] \cdot \frac{\delta^2 S_{\rm tot}}{\delta 
    \vf(\vec{x}) \, \delta \vf(\vec{y})} & = & 0
\end{eqnarray}
where:
\begin{eqnarray}
  T_{\eps}\left[ \vec{x},\vec{y},\vf \right] & = & T_{\eps}[\vf] - \eps^i (\vec{x}) \, 
    \frac{\partial}{\partial x^i} - \eps^i (\vec{y}) \, \frac{\partial}{\partial y^i} +  \\
&&~~~~~~~~~~~~~~~~~~~~+
    \lambda_{\eps}(\vec{x}) \left( 3 + \frac{\delta \beta(\vf(\vec{x}))}{\delta \vf(\vec{x})} \right)
 + \lambda_{\eps}(\vec{y}) \left( 3 + \frac{\delta \beta(\vf(\vec{y}))}{\delta \vf(\vec{y})} \right)\nonumber
\end{eqnarray}
This operator, acting the second functional derivative of $S_{\rm tot}$, satisfies: 
\begin{eqnarray}
 \left[\raisebox{12pt}{\,} T_{\eps_1}\left[ \vec{x},\vec{y},\vf \right],T_{\eps_2}\left[
   \vec{x},\vec{y},\vf \right] \right] & = & T_{\left[\eps_1,\eps_2\right]}\left[ 
   \vec{x},\vec{y},\vf  \right]
\end{eqnarray}
In other words it satisfies the Lie algebra associated with conformal transformations.

As a simple application of the conformal group, consider \eqref{tepsxy}
for constant $\lambda_{\eps}$ which implies $\eps^i = - \lambda_{\eps} \, x^i$. 
Taking $\chi_{f}(\vec{x})=0$ we find:
\begin{eqnarray}\label{posspaceCSeqn}
 \left( s^i \, \frac{\partial}{\partial s^i} + 6 + 2 \gamma +
\beta(\vf) \,  \frac{\partial}{\partial \vf} 
 \right) \,  S_{\rm tot}^{(2)}[a,\vf;\vec{s}] 
   & = & 0
\end{eqnarray}
where  $s^i = x^i - y^i$. This is the position-space version of the Callan-Symanzik equation
\eqref{1stCSeqn}. A similar computation demonstrates that 
$S_{\rm tot}^{(2)}[a,\vf;\vec{s}]$ is invariant under rotations.

\subsection{Inflation as a Holographic Quantum Field Theory}\label{subsec:inflationCFT}
The terminology we have introduced in this section to describe the
constraints due to diffeomorphism invariance is designed to suggest a formal 
analogy between the renormalized on-shell action for inflation and a 
three-dimensional Euclidean quantum field theory near a UV fixed point. 
The analogy is implemented by the map:
\begin{eqnarray}\label{duality}
	\Psi[\vf(\vec{x}), \tg_{ij}(\vec{x})] & = & Z[\vf(\vec{x}),\tg_{ij}(\vec{x})]
\end{eqnarray}
where the semi-classical wave function is constructed from the renormalized 
on-shell action $S_{\rm tot}$, with boundary data $\vf(\vec{x})$ 
and $\tg_{ij}(\vec{x})$: 
\begin{eqnarray}\label{wavefunctional}
  \Psi[\vf(\vec{x}),\tg_{ij}(\vec{x})] & = & \exp \left( i \,S_{\rm tot}
[\vf(\vec{x}),\tg_{ij}(\vec{x})]\raisebox{12pt}{\,}\right)
\end{eqnarray}
and the partition function refers to a Euclidean QFT with sources $\chi$ and 
$h_{ij}$:
\begin{eqnarray}\label{partitionfunction}
  Z[\vf(\vec{x}),\tg_{ij}(\vec{x})]
    & = & \left\langle  \exp \left( \int d^3 x \, \left[\frac{1}{2} \, 
          h_{ij}(\vec{x}) \,T^{ij}(\vec{x}) + \chi(\vec{x}) \, \OO(\vec{x})  \right] \right)
	  \right\rangle
\end{eqnarray}
The brackets in equation \eqref{partitionfunction} denote the QFT expectation value, with the background fields representing the couplings in the unperturbed theory. The semi-classical wave 
functional for inflation is then interpreted as the generating functional for correlators 
of the operator $\OO$ and the associated stress-energy tensor $T^{ij}$ in a 
quantum field theory. This hypothetical theory has been  displaced from the fixed point by an operator $\OO$ that is sourced by the boundary data $\vf(\vec{x},\tau_0)$. 

According to the 
identification \eqref{duality} we can calculate correlators of the QFT 
operators by functionally differentiating the partition function with respect to the 
sources, and then setting the sources equal to zero:
\begin{eqnarray}\label{correlatordefn}
   \langle \OO(\vec{x}_1) \ldots \OO(\vec{x}_n) \rangle & \equiv & 
     (-i)^n \, \left.\frac{\delta \Psi}{\delta \vf_{f}(\vec{x}_1) \ldots \delta
     \vf_{f}(\vec{x}_n)} \right\vert_{\chi= 0}= 
S_{\rm tot}^{(n)}[a,\vf;\vec{x}_1,\ldots, \vec{x}_{n}]
\end{eqnarray}
where the last equality utilized the definition \eqref{actionexpansion} of the 
coefficient $S_{\rm tot}^{(n)}$. An immediate consequence of this relation is that the 
one-point function $\langle\OO(\vec{x})\rangle$ vanishes, because there is no 
linear term in the expansion of the on-shell action. 
The differential equations for $S_{\rm tot}^{(n)}$ generated by bulk 
diffeomorphisms become, according to the analogy, equations that are familiar 
from renormalization theory. 
For example, the RG equation \eqref{rgeqn} for the background part of the 
on-shell action becomes an RG equation for the field theory partition function 
with no sources:
\begin{eqnarray}
 \left( a \, \frac{\partial}{\partial a} \, + \beta(\vf) \, \frac{\partial}{\partial \vf} \right)
 	\, Z[a,\vf] & = & 0
\end{eqnarray}
and the equation \eqref{1stCSeqn} that governs the coefficient of the  
quadratic term in the on-shell action becomes a Callan-Symanzik equation 
for the two-point function of the operator $\OO(\vec{x})$:
\begin{eqnarray}\label{CSeqn} 
   \left( a \, \frac{\partial}{\partial a} \, + \beta(\vf) \, \frac{\partial}{\partial \vf}
   	+ 2 \gamma(\vf) \right) \, \langle \OO(\vec{x}) \, \OO(\vec{y}) \rangle & = & 0
\end{eqnarray}

Strominger has proposed a relation similar to \eqref{duality} as a full-fledged quantum 
duality \cite{Strominger:2001pn, Strominger:2001gp}, known as the dS/CFT correspondence. 
This duality, if true, would relate a gravitational theory on (asymptotically) de Sitter 
spacetime to a (deformed) quantum conformal field theory on the boundary. 
Our approach is different both in its technical details 
(which follow Maldacena \cite{Maldacena:2002vr}) and in ambition. 
We regard \eqref{duality} as a convenient analogy which allow us to exploit much 
of our intuition about RG flows, because the physics of the inflating spacetime is 
governed by the same set of RG and Callan-Symanzik equations. However, 
similarities between the two systems does not, in and of itself, justify the idea of a 
true duality, and we do not appeal to a notion of a microscopically defined QFT 
dual to inflation.

\subsection{The Ward Identity}
Diffeomorphisms in the normal direction, which result in Weyl rescalings of the 
boundary data, constrain $S_{\rm tot}$ according to \eqref{functionalrgeqn}:
\begin{eqnarray}
 \int d^3 x \, \left[ 2 \, \tg_{ij} \, \frac{\delta S_{\rm tot}}{\delta \tg_{ij}} + 
   \beta(\vf) \, \frac{\delta S_{\rm tot}}{\delta \vf}\right] & = & 0
\end{eqnarray}
The map \eqref{duality} then gives a relation between the operator $\OO$ and the trace of the stress-energy tensor $T^{i}_{\,i}$:
\begin{eqnarray}\label{conformalWard}
	T^i_{\,i} + \beta \, \OO & = & 0
\end{eqnarray}
This is the conformal Ward identity for the field theory. For a conformally invariant theory the trace of the stress energy tensor vanishes\footnote{Inflation in four dimensions is analagous to a Euclidean field theory in three dimensions, so there is no conformal anomaly to consider.}. The operator $\OO$ displaces the field theory from its scale-invariant fixed point, and $T^i_{\,i}$ is no longer vanishing. Instead, it is proportional to the $\beta$-function of the operator  generating the flow away from the fixed point. 

The Conformal Ward Identity ensures that the terms involving scalars form gauge invariant combinations:
\begin{eqnarray}
  \frac{1}{2} \, h_{ij} T^{ij} + \chi \, \OO =  \psi \, T^{i}_{\,i} + \chi \, \OO
  =   \left( \chi-\beta\psi\right) \OO  
  & = & \chi_{f}(\vec{x}) \, \OO(\vec{x}) \\
  & = & - \psi_{\rm com}(\vec{x}) \, T^{i}_{\,i}(\vec{x})
\end{eqnarray}
This is because the $\beta$-function appearing in \eqref{conformalWard} is the same factor that relates the gauge invariant variables $\psi_{\rm com}$ and $\chi_{\rm flat}$:
\begin{eqnarray}
   \chi_{\rm flat}(\vec{x}) & = & - \frac{\vf\,'}{\HH} \, \psi_{\rm com}(\vec{x})
   			=  - \beta \, \psi_{\rm com}(\vec{x})
\end{eqnarray}
Thus, the scalar fluctuation in flat gauge, $\chi_{f}(\vec{x})$, sources the operator $\OO(\vec{x})$, and the comoving curvature perturbation $\psi_{\rm com}(\vec{x})$ sources the trace of the 
stress-energy tensor. We already used this when deriving \eqref{correlatordefn}.

The Ward Identity \eqref{conformalWard} holds as an operator equation, as opposed to a 
relation that applies to expectation values of the operators. This is because it is a 
consequence of the full master equation \eqref{mastereqn}, rather than the differential 
equations that constrain the coefficients of the Taylor expanded action. We can therefore
use it to relate higher correlators, such as:
\begin{eqnarray}\label{twopointfunctionidentity}
\langle T^i_{\,i}(\vec{x}) \, T^j_{\,j}(\vec{y}) \rangle & = & \beta(\vf)^2 \, \langle \OO(\vec{x}) \,\OO(\vec{y}) \rangle 
\end{eqnarray}
This is useful for the derivation of the CMB spectrum, presented in the next section. 

In addition to the conformal Ward identity we can obtain a Ward Identity associated 
with three dimensional diffeomorphisms. It follows from the momentum constraint \eqref{3Ddiffeoeqn}:
\begin{eqnarray}
	D^j \, T_{ij} + D_i \vf \, \OO & = & 0
\end{eqnarray}
Like the conformal Ward Identity this is an operator equation that holds outside of 
expectation values.

\section{Renormalization Group Improved CMB Power Spectrum}
\label{sec:IntegratingCS}
In this section we integrate the Callan-Symanzik equations and find the explicit
forms of the two point correlators which satisfy the constraints imposed diffeomorphism 
invariance. This leads to RG-improved versions of familiar results 
from slow-roll inflation. We consider scalar and tensor modes in turn.

\subsection{The Scalar Modes}
\label{subsec:PowerSpectrum}
Correlation functions of gauge invariant variables are easily computed
in terms of functional integrals weighted by the modulus-squared of 
the wave-functional $\Psi[\vf,\tilde{g}_{ij}]\sim e^{iS_{\rm tot}[\vf,\tilde{g}_{ij}]}$.
The simplest example is the two-point correlation function of the scalar field 
in flat gauge. Since the leading nontrivial terms in the action are quadratic:
\begin{eqnarray}
  S_{\rm tot}[a,\vf_{f}(\vec{x})] & = & S_{\rm tot}^{(0)} + \frac{1}{2} \, \int d^3 x \int d^3 y \, 
  	S_{\rm tot}^{(2)}[a,\vf,\vert\vec{x}-\vec{y}\vert\,] \, \chi_{f}(\vec{x}) \, 
\chi_{f}(\vec{y}) + 
	\ldots \\
& = & S_{\rm tot}^{(0)} + \frac{1}{2} \int d^3 k \, 
   \tilde S_{\rm tot}^{(2)}[a,\vf,k\,] \,  \, 
   \tilde \chi_{f}(\vec{k}) \, \tilde \chi_{f}(-\vec{k}) + \ldots
\end{eqnarray}
where Fourier modes are introduced as:
\begin{eqnarray} \label{Fouriermodes}
   \tilde \chi_{f}(\vec{k}) & = & \int d^3 x \, \frac{e^{i\vec{k}\cdot\vec{x}}}{(2\pi)^{3/2}} 
     \, \chi_{f}(\vec{x})
\end{eqnarray}
the functional integral is, to the leading order, a simple Gaussian. It  
gives \cite{Maldacena:2002vr}:
\begin{eqnarray}\label{chicorrelatorintegral}
  \langle \tilde{\chi}_{f}(\vec{k})\,\tilde{\chi}_{f}(-\vec{k})\rangle & = &  \, \int 
\DD\chi_{f}[\vec{q}] ~\, 
\chi_{f}(\vec{k})  \chi_{f}(-\vec{k}) \,\left|
  	\Psi[\chi_{f}(\vec{q})]\right|^{2} =  \frac{1}
{2{\rm Im}\tilde S_{\rm tot}^{(2)}[a,\vf,k\,] }
\end{eqnarray}
The Lorentzian path integral is interpreted as usual by continuation 
from Euclidean space. The ${\rm Im} S_{\rm tot}^{(2)}$ apprearing here is therefore
the (negative of) the real part of the Euclidean action which, according to the analogy 
between inflation and a QFT near a renormalization group fixed point, we can write as 
the two-point function $-{\rm Re}\langle\OO(\vec{k})\OO(\vec{-k})\rangle$.
The correlator can therefore be written as:
\begin{eqnarray}\label{chicorrelator}
  \langle \chi_{f}(\vec{k})\,\chi_{f}(-\vec{k})\rangle & = &  
 - \frac{1}{2{\rm Re}\langle \OO(\vec{k})\OO(\vec{-k})\rangle}
\end{eqnarray}
Note that the counterterms are real terms in the Lorentzian action, and so they 
do not contribute to the modulus of the wave function \cite{Maldacena:2002vr}.

As discussed in section \ref{sec:consequences}, diffeomorphism invariance
implies that the two-point correlation function of the operator ${\cal O}$  is
 constrained by the Callan-Symanzik equation \eqref{CSeqn}:
\begin{eqnarray}\label{realCS}
  \left( a\frac{\partial}{\partial a} + \beta(\vf) 
  \frac{\partial}{\partial \,\vf} + 2 \gamma(\vf) \right) \,
  \langle \OO(\vec{x})\,\OO(\vec{y})\rangle & = & 0
\end{eqnarray}
This equation can be solved by a procedure that is standard in renormalization group
theory.  We first use dimensional analysis on the three-dimensional boundary to write 
the correlator in the form:
\begin{eqnarray}
\label{ansF}
 \langle \OO(\vec{x})\,\OO(\vec{y})\rangle & = & \frac{1}{|\vec{x}-\vec{y}|^6}
F(|\vec{x}-\vec{y}|a,\vf)
\end{eqnarray}
To arrive at this we also appeal to rotational invariance and  
recall that $|\vec{x}-\vec{y}|a$ 
is the physical length, as opposed to the
coordinate length $|\vec{x}-\vec{y}|$. It is useful to Fourier transform the
position space Callan-Symanzik equation \eqref{posspaceCSeqn}:
\begin{eqnarray}\label{momentumCS}
  \left( 3 - \vec{k} \cdot \frac{\partial}{\partial \vec{k}} + \beta(\vf) 
  \frac{\partial}{\partial \,\vf} + 2 \gamma(\vf) \right) \,
  \langle \OO(\vec{k})\,\OO(-\vec{k})\rangle & = & 0
\end{eqnarray}
and the {\it ansatz} \eqref{ansF}:
\begin{eqnarray}
\label{kansF}
       \langle \OO(\vec{k})\,\OO(-\vec{k})\rangle & = & k^3 \, \tilde{F}\left(\frac{k}{a},\vf\right)
\end{eqnarray}
The ratio $k/a$ is the physical momentum. The next step in solving the 
Callan-Symanzik equation
is to introduce a ``running coupling" $\vfb(k/a, \vf)$ defined by:
\begin{eqnarray}
\label{runningdef}
	k \, \frac{\partial \vfb}{\partial \, k} & = &  \beta(\vfb) 
\end{eqnarray}
In the present context the ``running coupling" is just the ``rolling" scalar field 
of the background, evolving according to the FRW equations. 
The running coupling follows $\vf$ along the 
renormalization group flow back towards an arbitrary reference scale $M$,
with $\vfb(M,\vf)=\vf$. Employing \eqref{kansF} and \eqref{runningdef}
in solving \eqref{momentumCS}  we find that 
$\langle \OO(\vec{k})\,\OO(-\vec{k})\rangle$ must take the form:
\begin{eqnarray}
	\langle \OO(\vec{k})\,\OO(-\vec{k})\rangle & = & k^3 \, 
	\tilde{F}_{0}\left(\vfb\left(k/a,\vf\right)\right) \, \exp \left[  \int_{aM}^{k}  d 
	\log\left(\frac{k'}{a M}\right)~ \, 2\,\gamma(\vfb(k'/a,\vf))\right]
\end{eqnarray}
where $\tilde{F}_{0}$ is some unknown function.
In verifying this it is useful to note that the running coupling satisfies:
\begin{eqnarray}
\label{usebeta}
	\beta(\vf) \frac{\partial \vfb}{\partial \vf} & = &  \beta(\vfb) 
\end{eqnarray}
To summarize the computation thus far,
we have used the Callan-Symanzik equation to write the two-point 
correlator of the scalar field as:
\begin{eqnarray}
\label{toptcorr}
	\langle \tilde{\chi}_{f}(\vec{k})\,\tilde{\chi}_{f}(-\vec{k})\rangle & = & 
	\frac{1}{2k^3\tilde{F}_{0}\left(\vfb\left(k/a,\vf\right)\right) }\, \exp 
	\left[  -\int_{aM}^{k}  d \log\left(\frac{k'}{a M}\right)~ \, 2\,\gamma(\vfb(k'/a,\vf))\right]
\end{eqnarray}
The scalar field $\chi_f(x)$ has mass dimension $1$ in four dimensions; and so 
the function $\tilde{F}_{0}$ has mass dimension $-2$. In
order to extract this remaining dimensionful factor it is useful to introduce
a running Hubble constant $H(\vfb)$ which takes the form:
\begin{eqnarray}
  H(\vfb(k/a,\vf)) & = & H(\vf) \, \exp\left( - \frac{1}{2} \, \int_{aM}^{k} d 
    d \log\left(\frac{k'}{a M}\right) \, \beta(\vfb(k'/a,\vf))^{2}\right)
\end{eqnarray}
To verify this expression differentiate with respect to $\vf$ on both sides,
use the relations  \eqref{runningdef} \eqref{usebeta} to simplify the
derivatives, and the relation:
\begin{eqnarray}
	\beta(\vf) & = & -2\,\frac{\pvf H}{H}
\end{eqnarray}
which is easily derived from the FRW equations and the relation $H={\cal H}/a$
between the physical Hubble factor $H$ and the conformal Hubble factor ${\cal H}$.
We can now extract the dimensionful factor from the 
correlator \eqref{toptcorr} and write it as:
\begin{eqnarray}\label{integratecorrelator}
   \langle \tilde{\chi}_{\rm flat}(\vec{k})\,
\tilde{\chi}_{\rm flat}(-\vec{k})\rangle & = & \frac{H(\vf)^2}{2k^3}~ {\cal A}_s(\vfb(k/a,\vf)) 
    \exp\left( -\int_{aM}^{k} d \log\left(\frac{k'}{a M}\right) \left(\raisebox{12pt}{\,}
    \beta(\vfb(k'/a,\vf))^2 \right. \right.\nonumber 
    \\ 
    & & \hspace{120pt} + \left. \left. 2\,\gamma(\vfb(k'/a,\vf))\raisebox{12pt}{\,}
    \right)\raisebox{15pt}{\,}\right) 
\end{eqnarray}
where ${\cal A}_s$ is a dimensionless amplitude. 

In cosmology it is natural to compute correlation functions of the comoving curvature
perturbation $\psi_{\rm com}=-\beta(\vf)^{-1}\chi_{\rm flat}$, instead of the those
of the scalar field. This is because this quantity is purely geometrical, and so will not 
evolve while the perturbation is beyond the horizon\footnote{The comoving curvature perturbation is commonly denoted by either $\zeta$ or ${\cal R}$.}. The two-point 
correlation function of the comoving curvature perturbation is:
\begin{eqnarray}\label{psiOrelation}
  \langle \tilde \psi_{\rm com}(\vec{k}) \tilde \psi_{\rm com}(-\vec{k}) \rangle & = & 
\frac{1}{\beta(\vf)^2}\langle \tilde \chi_{\rm flat}(\vec{k}) \tilde \chi{\rm flat}(-\vec{k}) \rangle 
\end{eqnarray}
A conventional way to express two-point correlators in cosmology is to
introduce the power-spectrum for the scalar:
\begin{eqnarray}
\label{powerdef}
P_s(k) & = & \frac{k^3}{2 \pi^2} \,
\langle \tilde\psi_{\rm com}(\vec{k})\,\tilde\psi_{\rm com}(-\vec{k})\rangle
\end{eqnarray}
Collecting formulae, our result for the scalar power spectrum becomes:
\begin{eqnarray}\label{integratedcorrelator}
   P_s(k) & = &
\frac{H(\vf)^2}{(2\pi\beta(\vf))^2}~ {\cal A}_s(\vfb(k/a,\vf)) 
    \exp\left( -\int_{aM}^{k} d \log(\frac{k'}{aM}) \left(\raisebox{12pt}{\,}
    \beta(\vfb(k'/a,\vf))^2 \right. \right.\nonumber 
    \\ 
    & & \hspace{120pt} + \left. \left. 2\,\gamma(\vfb(k'/a,\vf))\raisebox{12pt}{\,}
    \right)\raisebox{15pt}{\,}\right) 
\end{eqnarray}

The amplitude ${\cal A}_s$ cannot be determined from diffeomorphisms alone;
it depends on dynamical content. Furthermore, ${\cal A}_s$ depends on the arbitrary 
scale $M$ in such a manner that the total correlator is independent of $M$. 
In the context of slow-roll inflation it can be shown that, for $M\gg H$, the 
amplitude ${\cal A}_s=1$. The content of the expression \eqref{integratedcorrelator} 
then is that it gives the correlator {\it all} $k$, given this 
dynamical input at $k=Ma$. Since the scale of main interest is $k=aH$, and the slow-roll 
computations are accurate well after horizon crossing $M\gg H$, the corrections 
we compute depend on the large ratio $M/H\gg 1$, and so they can be significant.
Our formula is thus a renormalization group improved version of the standard
computation. It would be interesting to compare our result to the more 
conventional ones for specific inflationary models. 

The scaling behavior of the power spectrum is characterized by the spectral index, 
defined by fitting to a powerlaw 
\begin{eqnarray}
\label{scalarinddef}
P_s(k)&\propto&\left( \frac{k}{aH}\right)^{n_s-1}
\end{eqnarray}
The 
scaling violations inherent in our result \eqref{integratedcorrelator} 
translates in the simplest instance into the spectral index:
\begin{eqnarray}\label{spectralindex}
  n_{s}-1 & = & k\frac{\partial}{\partial \, k} \, \log{P_s(k)}
 =  - \beta(\vf)^2 - 2\,\gamma(\vf)
\end{eqnarray}
To compare this result with the standard slow-roll inflation results, we can relate
the RG parameters $\beta$ and $\gamma$ to the usual slow-roll expansion parameters.
\begin{eqnarray}
\beta({\vf})^2 &=&  2\,\bar{\epsilon} \,\,= \,\,  4\,\left( \frac{\partial_\vf H}{H}\right)^2 
\label{epsilondef} \\
\gamma({\vf}) &= & \bar{\epsilon}-\bar{\eta} \,\, = \,\, 2\,\left( \frac{\partial_\vf   
   H}{H}\right)^2  -2\,\frac{\partial^2_\vf H}{H}
\label{etadef}
\end{eqnarray}
With these identifications our result \eqref{spectralindex} is seen to agree
with:
\begin{eqnarray}\label{standardns}
n_s & = & 1 - 4\,\bar{\epsilon} + 2\,\bar{\eta}
\end{eqnarray}
which is the standard first order result for slow-roll inflation. 

In obtaining \eqref{spectralindex} we computed the derivative at $k=M$ and neglected 
the $k$-dependence implicit in ${\cal A}_s$. This is justified only to the leading order, 
{\it i.e.} when the spectral index is considered a constant. An improved result
can be obtained by taking into account the evolution of the spectral index with $k$.
The result obtained this way will not be exact, since we do not in general know 
${\cal A}_s$, but it will be more accurate than the standard results, since our 
treatment is RG-improved: we resum the large logarithms from terms that
are formally of higher order.

\subsection{The Tensor Modes}
Up to this point we have neglected contributions due to the tensor modes, which are 
completely decoupled from the scalar fluctuation at quadratic order. We will now
repeat the analysis with the tensor modes included. In the analogy with a three 
dimensional Euclidean field theory, the starting point is the full partition function,
with the tensor modes included. The couplings between sources and operators are of the 
form:
\begin{eqnarray}
\int d^3 x \, \left[ \frac{1}{2} \, h_{ij} \, T^{ij} + \chi \, \OO \right] 	
& = &  \int d^3 x \, \left[ \frac{1}{2} \,\phi_{ij} \, T^{i}_{\,j}  +  \chi _{f} \, \OO \right]
\end{eqnarray}
For the scalars we used the conformal Ward identity \eqref{conformalWard} to rewrite 
$\psi T^i_{\,i}+\chi{\cal O}$ as the gauge invariant quantity $\chi_f{\cal O}$.
If we choose a basis $e^{\pm}_{ij}$ for the tensor modes:
\begin{eqnarray}
\label{tensorbasis}
  \phi_{ij} & = & 2 \left(  u_+ \, e^+_{ij} + u_- \, e^-_{ij} \right) 
\end{eqnarray}
with $e^{\pm}_{ij} e^{\pm}_{ij}  =  1$ and $e^{\pm}_{ij} e^{\mp}_{ij}  =  0$, then the two 
polarization states of the tensor appear in the action \eqref{tensorAction} 
as minimally coupled, massless scalars with a canonical kinetic term. In the partition 
function they serve as sources for operators $t_+$ and $t_-$, corresponding to the two components of the traceless, divergence-free part of the field theory stress tensor:
\begin{eqnarray}
  Z[\vf(\vec{x}), \tg_{ij}(\vec{x})] & = & \left\langle  \exp \left( \int d^3 x \left[ u_+(\vec{x}) t_+(\vec{x}) + u_-(\vec{x}) t_-(\vec{x})
  	+ \chi_{f}(\vec{x}) \, \OO(\vec{x}) \right]\right) \right\rangle
\end{eqnarray}
To quadratic order we can compute the Gaussian integral as in 
\eqref {chicorrelatorintegral} and relate the two-point function for each mode 
$u_{\pm}$ to the correlator of the corresponding operator $t_{\pm}$:
\begin{eqnarray}\label{utrelation}
 \langle \tilde{u}_+(\vec{k})\tilde{u}_+(-\vec{k})\rangle & = & 
\frac{1}{2 \, \langle \tilde{t}_+(\vec{k}) \tilde{t}_+(-\vec{k}) \rangle}
\end{eqnarray}

The Callan-Symanzik equation for correlators of the tensor modes is derived
from \eqref{masterweyl} by varying with respect to the tensor sources. 
Since the $\beta$-function is independent of the gravitational field ---
it depends on the scalar field only --- the variations commute with all
terms in the equation. The Callan-Symanzik equation of the
tensor modes is therefore simply:
\begin{eqnarray}
  \left( a\,\frac{\partial}{\partial \,a} + \beta(\vf)\,\frac{\partial}{\partial \, \vf}\right)
   \, \langle t_{\pm}(\vec{k}) \, t_{\pm}(-\vec{k})\rangle & = & 0
\end{eqnarray}
with no term corresponding to an anomalous dimension. 
It is simple to integrate this equation by repeating the steps that lead to 
\eqref{integratecorrelator} for the correlator of scalar fields, with
the simplification that now the anomalous dimension $\gamma_t=0$. 
The correlator of the tensor fields $\tilde{u}_{\pm}$ thus takes the form:
\begin{eqnarray}\label{integratedtensorcorrelator}
  \langle \tilde{u}_{\pm}(\vec{k}) \, \tilde{u}_{\pm}(-\vec{k}) \rangle & = & 
\frac{H(\vf)^2}{2\,k^3} {\cal A}_t(\vfb(k/a,\vf))    
\, \exp\left(-\int_{aM}^{k} \bns d \log\left({\frac{k'}{aM}}\right) \, \beta(\vfb(k'/a,\vf))^2\right)
\end{eqnarray}
The power spectrum for the tensor modes is introduced as:
\begin{eqnarray}
  P_t(\vec{k}) & = & \frac{k^3}{2\pi^2} \, \langle \tilde{\phi}_{ij}(\vec{k}) \, \tilde{\phi}_{ij}(-\vec{k}) 
  \rangle \\
   & = & \frac{k^3}{2\pi^2} \, 8 \, \langle \tilde{u}_+(\vec{k}) \, 
\tilde{u}_+(-\vec{k})\rangle
\end{eqnarray}
Factors of two appear in the second line from the decomposition \eqref{tensorbasis}
and also because we have written the 
result entirely in terms of $\langle u_+ u_+ \rangle$; the $u_+$ and $u_-$ 
correlators are identical in the absence of polarizing sources.
Our result for the RG improved power spectrum of the tensor modes becomes:
\begin{eqnarray}\label{tensorpower}
 P_t(\vec{k})  & = & 
\frac{H(\vf)^2}{(2\pi)^2}~ 8{\cal A}_t(\vfb(k/a,\vf))    
\, \exp\left(-\int_{aM}^{k} d \log({\frac{k'}{aM}}) \, \beta(\vfb(k'/a,\vf))^2\right)
\end{eqnarray}
As for scalars, the amplitude ${\cal A}_t$ is not determined by symmetries alone.
However, in slowroll inflation, ${\cal A}_t=1$ well after horizon crossing. Our 
expression then gives the full dependence on the scale $k$, and it acts as
a resummation of the corrections to slow roll which, in general, can be 
large.

The spectral index $n_t$ for the tensor modes is defined by fitting the
power spectrum to the form:
\begin{eqnarray}
\label{tensorinddef}
P_t(\vec{k})&\propto&\left( \frac{k}{aH}\right)^{n_t}
\end{eqnarray}
Note that the convention for the tensor index $n_t$ is shifted by one in 
comparison with the scaler index \eqref{scalarinddef}; so 
scale invariance corresponds to $n_t=0$ for tensor modes. 
Repeating the computation of the spectral index for scalar 
modes \eqref{spectralindex} we find: 
\begin{eqnarray}
 n_t & = & k\frac{\partial}{\partial \, k} \, \log{P_t(\vec{k})}
 =  - \beta(\vf)^2 
\end{eqnarray}
for the spectral index of the tensor modes. Our result agrees with the
standard result $n_t = -2\epsilon$, when it is expressed in in terms of the 
slow-roll parameter $\epsilon$ \eqref{epsilondef}.

The ratio of the tensor and power spectra is:
\begin{eqnarray}
\frac{P_t}{P_s} &=& {8\beta(\vf)^2}  \frac{{\cal A}_t(\vfb(k'/a,\vf))}
{{\cal A}_s(\vfb(k'/a,\vf))}
\exp\left(\int_{aM}^{k} d \log({\frac{k'}{aM}}) \, 2\gamma(\vfb(k'/a,\vf))\right)
\end{eqnarray}
In slow roll inflation where ${\cal A}_t={\cal A}_s=1$ and the exponential
is negligible this reduces to the famous consistency condition:
\begin{eqnarray}
\frac{P_t}{P_s} &=& {8\beta(\vf)^2} = -8n_t
\end{eqnarray}
on the physical observables. Our modest elaboration of this standard
result is that our result incorporates the evolution of this ratio with scale $k$. 

\section{Outlook}
Before concluding this paper we would like to briefly mention several applications 
and extensions of our results which we think are worth pursuing further:
\begin{itemize}
\item
A concrete result of the investigation in this paper are the explicit, gauge invariant 
actions \eqref{tensorAction}, \eqref{psicomAction}, and \eqref{MukhanovAction} for 
fluctuations in spacetimes with a boundary. There are numerous applications of actions 
such as these in the context of the AdS/CFT correspondence generally, and specificaly 
in the context of warped brane-world models, such as those of Randall-Sundrum type. 
Moreover, the nontrivial role of counterterms in implementing diffeomorphism invariance,
and so obtaining gauge invariant actions, might well resolve various puzzles
in those other settings. 
\item
We have introduced our concept of RG improved computations of CMB spectra,
based on integrating the constraints from invariance under infinitesimal diffeomorphisms, 
as expressed by the Callan-Symanzik equation. It would be interesting to make
this procedure explicit for concrete potentials of interest in inflationary cosmology,
and so identify cases where these corrections are in fact significant. 
\item
It would be interesting to extend our analysis beyond quadratic order, and obtain constraints
on correlators of more fields than two. The main difficulty in carrying this out is that
one must implement diffeomorphism invariance, or gauge invariance, beyond the 
leading order. This is nontrivial, although the Hamiltonian formalism should automatize
a significant part of the work. 

\item
The cancellation of divergent terms in the on-shell action by a finite number of
boundary counterterms is dependent on some constraints on the backgrounds:
it is known to work for asymptotically (A)dS-spacetimes, and we show that the mechanism
extends to large classes of FRW cosmologies. The situation in more general
spacetimes, including asymptotically flat ones, is less clear. It would be interesting
to explore whether our method is useful for adapting 
the boundary counterterm method to a wider class of spacetimes. 
Additionally, the method may serve as a basis for constraining the boundary terms which 
make the notion of quasi-local mass and energy in gravitational theories inherently 
ambiguous \cite{Brown:2000dz,Balasubramanian:zh, Balasubramanian:2001nb}.

\end{itemize}

\section*{Acknowledgements}
We thank
V. Balasubramanian,
J. Distler,
D. Minic,
A. Strominger,
S. Weinberg,
and
particularly
J. Maldacena
for discussions.
Both of us are grateful for hospitality at the KITP (Santa Barbara) during phases of this work. 
This research was supported in part by the DoE through Grant No. 
DE-FG02-95ER40899 (Michigan) and in part by the National 
Science Foundation through Grant No. PHY99-07949 (Santa Barbara).

\pagebreak

\end{document}